\documentclass[preprint,12pt]{elsarticle}
\usepackage{graphicx}
\usepackage{amssymb}

\def\rt{\mathfrak{r}}

\newcommand{\rmd}{\ensuremath{\mathrm{d}}}
\newcommand{\rmi}{\ensuremath{\mathrm{i}}}
\newcommand{\rme}{\ensuremath{\mathrm{e}}}
\journal{Nuclear Physics B}
\begin{document}
\begin{frontmatter}

\title{An efficient method for computing genus expansions and counting numbers in the Hermitian matrix model}

\author[ucm]{Gabriel \'Alvarez\corref{cor1}}
\ead{galvarez@fis.ucm.es}

\author[ucm]{Luis Mart\'{\i}nez Alonso}
\ead{luism@fis.ucm.es}

\cortext[cor1]{Corresponding author}

\author[uca]{Elena Medina}
\ead{elena.medina@uca.es}

\address[ucm]{Departamento de F\'{\i}sica Te\'orica II,
                               Facultad de Ciencias F\'{\i}sicas,
                               Universidad Complutense,
                               28040 Madrid, Spain}

\address[uca]{Departamento de Matem\'aticas,
                      Facultad de Ciencias,
                      Universidad de C\'adiz,
                      11510 Puerto Real, Spain}

\begin{abstract}
We present a method  to compute the genus expansion of  the free energy 
of Hermitian matrix models from the large $N$ expansion of the recurrence coefficients
of the associated family of orthogonal polynomials. The method is based on the Bleher-Its deformation
of the model, on its associated integral representation of the free energy, and on a method for solving the
string equation which uses the resolvent of the Lax operator of the underlying Toda hierarchy.
As a byproduct we obtain an efficient algorithm to compute generating functions for the enumeration of
labeled $k$-maps which does not require the explicit expressions of the coefficients of the topological
expansion. Finally we discuss the regularization of singular one-cut models within this approach.
\end{abstract}


\end{frontmatter}
\section{Introduction}
In this paper we consider the ensemble of random Hermitian matrices
\begin{equation}
  \label{0.1}
  Z_{N}(\mathbf{g})
  =
  \int_{\mathbf{R}^N}
  \exp\left(-N\sum_{i=1}^N V(x_i,\mathbf{g})\right)
  \prod_{i<j}(x_i-x_j)^2 \rmd x_1\cdots \rmd x_N,
\end{equation}
for a given polynomial potential
\begin{equation}
  V(z,\mathbf{g}) = \sum_{n=1}^{2p} g_{n} z^{n}
\end{equation}
of degree $2p$ with real coefficients $\mathbf{g}=(g_1,\ldots,g_{2p})$ such that $g_{2p}> 0$
(we will not make explicit the dependence on $\mathbf{g}$ of the functions associated
with the model~(\ref{0.1}) unless there is risk of ambiguity).  For more than thirty
years~\cite{BR78,BL05,BL08,BL10,DI95,DI06} the asymptotic behavior of  the free energy
\begin{equation}
  \label{fre}
  F_N = -\frac{1}{N^2}\ln Z_{N}
\end{equation}
as $N\rightarrow \infty$ and its relation to the counting of Feynman graphs have been subjects of
intensive research. However, rigorous proofs of the existence of an asymptotic expansion
of $F_N$ in powers of $N^{-2}$ were provided only rather more recently
by Ercolani and McLaughlin~\cite{ER03} and by Bleher and Its~\cite{BL05}.
These analyses prove the existence of a  genus expansion of the form
\begin{equation}
  \label{bif20}
  F_{N}(\mathbf{g}) - F_N^{\mathrm{G}} \sim \sum_{k\geq 0} F^{(k)}(\mathbf{g}) N^{-2k},
\end{equation}
where $F_N^{\mathrm{G}}$ stands for the Gaussian free energy,
\begin{equation}
  F_N^{\mathrm{G}} = -\frac{1}{N^2}
                                     \ln\left( \frac{2\pi^{N/2}}{(2N)^{N^2/2}}\prod_{n=1}^N n!\right),
\end{equation}
under the assumption that there is a path $\mathbf{g}(t)$
in the space of coupling parameters connecting $V(z,\mathbf{g})$ to the Gaussian potential $z^2$
in such a way that $V(z,\mathbf{g}(t))$ is a regular one-cut model for all $t$.
The functions $F^{(k)}(\mathbf{g})$ are important objects because the
coefficients of their Taylor expansions at the Gaussian point $g_k^{\mathrm{G}}=\delta_{k,2}$
are generating functions for the enumeration of labeled $k$-maps with vertices involving
valences $1,\ldots,2 p$, where $2 p$ is the number of nonvanishing coupling parameters $g_n$. 
The general aim of the present work is the characterization of the structure of these
functions $F^{(k)}(\mathbf{g})$. (Incidentally, for multi-cut models it has been
shown~\cite{BO00,EY09} that in general the free energy exhibits an oscillatory behavior
as a function of $N$, and consequently topological expansions cannot exist.) 

The large $N$ asymptotics of the matrix model~(\ref{0.1}) is intimately connected with
the asymptotics of  the recurrence coefficients $r_{n,N}$ and $s_{n,N}$ in
the three-term recursion relation
\begin{equation}
  \label{rec}
  x P_{n,N}(x) = P_{n+1,N}(x)+s_{n,N} P_{n,N}(x) + r_{n,N} P_{n-1,N}(x),
\end{equation}
for the orthogonal polynomials $P_{n,N}(x)=x^n+a_{n-1}x^{n-1}+\cdots$
with respect to the exponential weight
\begin{equation}
  \label{pol}
  \int_{-\infty}^{\infty}P_{k,N}(x) P_{l,N}(x) e^{-N V(x)} \rmd x
  =
  \delta_{k,l} h_{k,N}.
\end{equation}
In particular~\cite{DE99b,SA97}, in the limit
\begin{equation}
  \label{lim}
   n\rightarrow \infty,\quad N\rightarrow \infty,\quad \frac{n}{N}\rightarrow T,
\end{equation}
the density of zeros of  $P_{n,N}(x)$ reduces to the  eigenvalue density
of the matrix model $Z_N(\mathbf{g}/T)$, and if $Z_N(\mathbf{g}/T)$ is a regular
one-cut model  then~\cite{BL08,DE99} the recurrence coefficients $r_{n,N}$ and
$s_{n,N}$ can be expanded in powers of $N^{-2}$~\cite{BL05}.

The methods that exploit this relation with orthogonal polynomials to calculate the asymptotics
of the free energy essentially consist of three steps:
\begin{enumerate}
  \item Determine  the free energy in terms of the recurrence coefficients.
  \item Obtain the asymptotic expansion of the recurrence coefficients.
  \item Use 1 and 2 to obtain the asymptotic expansion of the free energy.
\end{enumerate}
There are alternative methods based on solving the Ward identities for the partition function
(loop identities)~\cite{AM93}  (cf.~also~\cite{EY04}) or on formulating the matrix model as a conformal field
theory~\cite{KO10}. However, the structure of the
expressions for the $F^{(k)}(\mathbf{g})$ that these methods provide is
less suitable than ours to compute generating functions for the enumeration of labeled $k$-maps. 

In this paper we present an efficient method to calculate the topological expansion
of the free energy and to characterize its coefficients.  For simplicity we restrict our analysis to
Hermitian models associated to  even potentials $V(\lambda)$, where $\lambda=z^2$. 
We now preview how our approach performs the three steps and point out the differences with
respect to other schemes.

The classical method of Bessis, Itzykson and
Zuber~\cite{DI06, BE79, BE80, DE90, SH95, SH96, MA08} is based on the following identity (step~1):
\begin{equation}
  \label{rel1}
  F_N = -\frac{1}{N^2} \ln (N!)-\frac{1}{N}
              \left( \ln h_{0,N} + \sum_{n=1}^{N-1} \left(1-\frac{n}{N}\right) \ln r_{n,N} \right),
\end{equation}
wherein the expansion of $r_{n,N}$ is substituted (step 2), and the asymptotic behavior of $F_N$
is obtained by means of the Euler-Maclaurin summation formula (step 3).  However, some objections
to this approach were raised by Ercolani and McLaughlin (cf.~subsection~1.5 in~\cite{ER03}) due to
the use of the asymptotic series of $r_{n,N}$ as a uniform expansion valid even for $n=1$ as
$N\rightarrow\infty$. This objection triggered the interest in alternative strategies for
step~1~\cite{BL05,ER03,ER08}. For example,  Ercolani, McLaughlin and Pierce~\cite{ER08} derived a
hierarchy of second order differential equations to determine the coefficients of the  expansion~(\ref{bif20})
from those of the the asymptotic expansion of $r_{n,N}$. 
In our work we  use instead  the  Bleher-Its  integral representation \cite{BL05}
\begin{equation}
  \label{bif}
  F_{N}(\mathbf{g})
  =
  F_N^{\mathrm{G}}
 +
  \int_1^{\infty} \frac{1-t}{t^2}
  \left[r_{N,N}({\mathbf{g}(t)}) \left( r_{N-1,N}({\mathbf{g}(t)})
                                                          + r_{N+1,N}({\mathbf{g}(t)})\right)-\frac{1}{2}\right]\rmd t.
\end{equation}
where $\mathbf{g}(t)$ denotes  the Bleher-Its deformation~\cite{BL05}
\begin{equation}
  \label{def}
   V(\lambda,\mathbf{g}(t)) = (1-1/t) \lambda + V(\lambda/t,\mathbf{g}),\quad 1\leq t<\infty,
\end{equation}
or explicitly in terms of the coupling parameters,
\begin{equation}
  g_2(t) = 1 - \frac{1}{t} + \frac{g_{2}}{t}, \qquad  g_{2k}(t)=\frac{g_{2k}}{t^k},\quad k\geq 2 .
\end{equation}

For step~2 the standard methods~\cite{BL05,DI95} use recursive equations for the
coefficients $r_{n,N}$ (string equations) to determine expansions of the form
\begin{equation}
  \label{in1}
  r_{n,N}(\mathbf{g}) \sim \sum_{k\geq 0} r_k(T,\mathbf{g})\epsilon^{2k},\quad \epsilon=\frac{1}{N}.
\end{equation}
Bessis, Itzykson and Zuber~\cite{BE80,IT80} formulated the string equation in terms
of summations of paths over a certain staircase. Some years later  Shirokura~\cite{SH95,SH96}
developed a general method to perform these summations and characterize the coefficients
of the expansion~(\ref{in1}) in terms of the single function
\begin{equation}
  \label{in2}
  W(r_0,\mathbf{g})= \sum_{n=1}^p {2 n \choose n} n g_{2n} r_0^n.
\end{equation}
To solve the string equation, in this paper we introduce a generating function $U_{n,N}(\lambda)$
associated to the resolvent of the finite-difference Lax operator $L$ of the underlying Toda
hierarchy~\cite{GE91}
\begin{equation}
  \label{int3}
  L P_{n,N} = P_{n+1,N} + r_{n,N} P_{n-1,N},
\end{equation}
 and determine the coefficients of~(\ref{in1}) from the function~(\ref{in2}) as rational functions
of the leading coefficient $r_0$.  This approach is simpler that Shirokura's method, is particularly suitable
for symbolic computation, can be applied to a generic potential, and permits an easy characterization
of the asymptotics of the Bleher-Its deformation $ r_{n,N}(\mathbf{g}(t))$ of  the recurrence coefficient. 

Finally, regarding step~3,  the Bleher-Its representation~(\ref{bif}) allows us to express the coefficients
of the topological expansion as
\begin{equation}
  \label{int4}
  F^{(k)}(\mathbf{g}) = \int_0^{r_0} R_k(\xi,\mathbf{g})\rmd\xi,\quad k\geq 1,
\end{equation}
where the integrands $R_k(\xi,\mathbf{g})$ are rational functions of $\xi$ which can be computed
explicitly in terms of the function~(\ref{in2}). 

The layout of this paper is as follows. In section~2 we briefly review the basic facts about
matrix models which are required to discuss the Bleher-Its deformation and the conditions that
ensure the existence of the expansion~(\ref{bif20}). Then we apply our results to the quartic model
and to the sixtic model of Brezin, Marinari and Parisi~\cite{BR90}. Section~3 is devoted to the asymptotic
expansion~(\ref{in1})  of the recurrence coefficient for general models $V(\lambda,\mathbf{g})$
and their respective Bleher-Its deformations $V(\lambda,\mathbf{g}(t))$. In particular, we rederive
in a much shorter way the expressions for the coefficients $r_1(T,\mathbf{g})$ and $r_2(T,\mathbf{g})$ 
found by Shirokura~\cite{SH95,SH96}.  Section~4 deals with the asymptotics of the free energy.
From the Bleher-Its representation~(\ref{bif}) and the expansion of the recurrence coefficients we
obtain the integral expression~(\ref{int4}). We evaluate explicitly the integrals for the 
$F^{(k)}$  up to genus~3 in the general case and find, except for a coefficient in the expression of
$F^{(3)}$, the same results found by Shirokura~\cite{SH95,SH96}. We also check that the
expressions of these coefficients for general two-valence models reduce to those obtained using
the Ercolani-McLaughlin-Pierce method~\cite{ER08}. In the brief section~5 we discuss how to apply
our method  to compute counting maps functions and present explicit calculations for two and three valence models.
In section~6 we formulate a ``triple scaling'' method to regularize the free energy expansion of a
class of singular models (the singular one-cut case) and show how the Painlev\'e~I hierarchy
emerges in our approach. The paper ends with a brief summary.
\section{The Bleher-Its deformation}
To  calculate the asymptotic behavior of $F_N(\mathbf{g})$ as $N\to\infty$ using the Bleher-Its formula~(\ref{bif})  we need  the
asymptotics of the deformed recurrence coefficients $r_{n,N}({\mathbf{g}(t)})$ where $\mathbf{g}(t)$ is the
Bleher-Its deformation of  $\mathbf{g}$. In this section we study the action of this deformation  on the space of
coupling parameters.
As relevant examples  we analyze the Bleher-Its deformation for models associated to quartic potentials
and to the sixtic potentials of Brezin, Marinari and Parisi~\cite{ BR90}.

The continuum limit of $r_{n,N}(\mathbf{g}(t))$ depends on the number $q$ of cuts of the model $V(\lambda,\mathbf{g}(t))/T$.
In appendix~A we summarize the method to determine the number of cuts of a generic hermitian model.
The endpoints of a $q$-cut eigenvalue support $J=\cup_{j=1}^q (\alpha_j,\beta_j)$ are the solutions of the system
of $2 q$ equations~(\ref{e1})--(\ref{e3}) but, in general, several such systems of equations corresponding
to different values of $q$ may have admissible solutions for one and the same model. Among these candidate solutions,
the correct value of $q$ is uniquely determined by the additional set of inequalities~(\ref{des1})--(\ref{des3})
on the polynomial $h(z)$ defined by
\begin{equation}
  \label{0.3}
  \frac{V_z(z)}{w_1(z)} = h(z)+\mathcal{O}(z^{-1})\quad\mbox{as }z\rightarrow \infty,
\end{equation}
where $w_1(z)$ is the branch of the function
\begin{equation}
  w(z) = \sqrt{\prod_{i=1}^{q} (z-\alpha_i)(z-\beta_i)}
\end{equation}
with asymptotic behavior $w_1(z)\sim z^q$ as $z\rightarrow\infty$.
In turn, the polynomial $h(z)$ is related to the eigenvalue density $\rho(x)$ by
\begin{equation}
  \label{0.2}
  \rho(x) = \frac{h(x)}{2\pi\rmi} w_{1,+}(x)\quad\mbox{for }x\in J,
\end{equation}
where $w_{1,+}(x)$ denotes the boundary value of $w_1(z)$ on $J$ from above. 

We restrict our considerations to even potentials of the form
\begin{equation}
  \label{eve}
  V(\lambda,\mathbf{g}) = \sum_{j=1}^p g_{2j} \lambda^{j},
  \quad
  \lambda = z^2,
  \quad g_{2p}>0,
\end{equation}
where the  coupling constants $\mathbf{g}=(g_2,g_4,\ldots,g_{2p})$ run on a certain region $G$ of  $\mathbf{R}^p$.
The \emph{phase diagram} of the corresponding family of matrix models is introduced through  the decomposition
\begin{equation}
  \label{cou}
  G = \bigcup_{q=1}^p \overline{G}_q,
\end{equation}
where $\mathbf{g}\in G_q$ if and only if $\mathbf{g}$  determines a $q$-cut regular model (cf.~appendix~A).
We will refer to  $G_q$ as the $q$-cut phase of the family~(\ref{eve}) of Hermitian models.
For even potentials the eigenvalue support $J$ is symmetric with respect to the origin, and in the
one-cut case the endpoints of $J=(-\alpha,\alpha)$ are determined by the single equation~(\ref{e3}), which in terms
of $V$ reduces to
\begin{equation}
  \label{q1}
  \oint_{\gamma} \frac{\rmd \lambda}{2\pi\rmi} V_{\lambda}(\lambda) \sqrt{\frac{\lambda}{\lambda-\alpha^2}} =1.
\end{equation}

Regarding the behavior of a particular model $\mathbf{g}\in G$ with respect to its Bleher-Its deformation,
we will consider two cases in our analysis:  the \emph{regular one-cut case}, in which ${\mathbf{g}(t)}\in G_1$
for all $t\geq 1$, and the \emph{singular one-cut case} in which ${\mathbf{g}(t)}\in G_1$ for $t>1$ but 
$\mathbf{g}=\mathbf{g}(1)$ determines a singular model (cf.~appendix~A).
\subsection{The quartic model\label{sec:qm}}
The quartic model
\begin{equation}
  \label{bip}
  V(\lambda,\mathbf{g}) = g_2\lambda + g_4\lambda^2
\end{equation}
in the region
\begin{equation}
  G = \{\mathbf{g}=(g_2,g_4)\in \mathbf{R}^2: g_4>0 \},
\end{equation}
only exhibits $q=1$ and $q=2$ phases~\cite{BL99}.  The $q=1$ phase $G_1$ can be written as the union
\begin{equation}
  \label{ph1}
  G_1=G_1^{(1)} \cup G_1^{(2)},
\end{equation}
where 
\begin{eqnarray}
  G_1^{(1)} & = & \{ (g_2,g_4)\in \mathbf{R}^2: g_2\geq 0, g_4>0\},\\
  G_1^{(2)} & = & \{ (g_2,g_4)\in \mathbf{R}^2: g_2<0, g_4>0, g_2>-2\sqrt{g_4}\},
\end{eqnarray}
and the $q=2$ phase $G_2$ is given by
\begin{equation}
  G_2=\{ (g_2,g_4)\in \mathbf{R}^2: g_4>0, g_2<-2 \sqrt{g_4} \}.
\end{equation}
The phase diagram features the critical curve
\begin{equation}
  \label{eq:cc}
  g_2=-2 \sqrt{g_4},
\end{equation}
which demarcates the transition line between the two phases.

\begin{figure}
  \begin{center}
  \includegraphics[width=10cm]{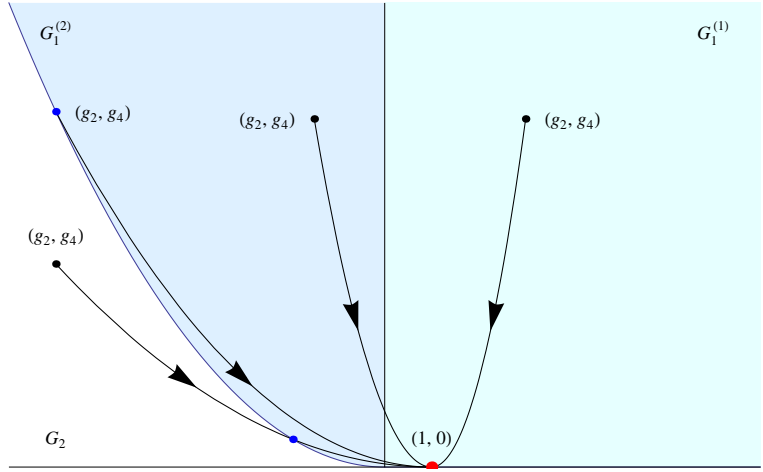}
  \end{center}
  \caption{Deformation paths ${\mathbf{g}(t)}$ for the quartic potential. From left to right: a deformation of a two-cut regular model,
  a deformation in the singular one-cut case and two deformations in the regular one-cut case.\label{fig:tau}}
\end{figure}

Consider now the Bleher-Its deformation (cf.~figure~\ref{fig:tau})
\begin{equation}
  {\mathbf{g}(t)} = (g_2(t),g_4(t)) = \left(\frac{t-1+g_2}{t}, \frac{g_4}{t^2}\right).
\end{equation}
If $\mathbf{g}\in G_1^{(1)}$  then $g_2>0$ and  $g_2(t)>0$ for all $t>1$.
Therefore  ${\mathbf{g}(t)} \in G_1$  for all $t>1$. If $\mathbf{g} \in G_1^{(2)}$, then
$g_2>-2\sqrt{g_4}$ and
\begin{equation}
  g_2(t) = \frac{t-1+g_2}{t}>-2 \frac{\sqrt{g_4}}{t}=-2\sqrt{g_4(t)}.
\end{equation}
Hence if $\mathbf{g} \in G_1^{(2)}$ we also have that ${\mathbf{g}(t)}\in G_1$ for all $t>1$.
On the other hand, it is elementary to see that if $\mathbf{g} \in G_2$ or is on the
critical curve~(\ref{eq:cc}) then ${\mathbf{g}(t_0)}$
is on the critical curve for $t_0=1-g_2-2 \sqrt{g_4}$. Summing up,
\begin{enumerate}
  \item If $\mathbf{g}\in G_1$ then ${\mathbf{g}(t)}\in G_1$ for all $t\geq1$.
  \item If $\mathbf{g} \in G_2$ then ${\mathbf{g}(t)}$ crosses the critical curve at $t_0=1-g_2-2 \sqrt{g_4}$.
  \item If $\mathbf{g}$ is on the critical curve~(\ref{eq:cc}) then ${\mathbf{g}(t)}\in G_1$ for all $t>1$.
\end{enumerate}
\subsection{The  Brezin-Marinari-Parisi model}
In~\cite{BR90} Brezin, Marinari and Parisi considered the potentials $V(z)/T$ with
\begin{equation}
  \label{bmpp}
  V(z) = 90 z^2 - 15 z^4 + z^6
\end{equation}
to generate a non-perturbative ambiguity-free solution of a string model. The models
$V(z)/T$ define a path in the space of coupling constants
\begin{equation}
G=\{\mathbf{g} =(g_2,g_4,g_6)\in \mathbf{R}^3: g_2>0, g_4<0, g_6>0 \}
\end{equation}
of the family of even sixtic potentials
\begin{equation}
  \label{eq:v6}
  V(\lambda,\mathbf{g}) = g_2 \lambda + g_4 \lambda^2 + g_6 \lambda^3,\quad (\lambda=z^2).
\end{equation}
We refer to appendix~A for the proof of the following facts: (i) the inequality
\begin{equation}
  \label{eq:in}
  \frac{5 g_2 g_6}{2 g_4^2} > 1
\end{equation}
determines an open subset of the one-cut phase $G_1$;
(ii) the boundary $\Gamma$ of this subset is the elliptic cone
\begin{equation}
  \label{eq:ec}
  5 g_2 g_6 = 2 g_4^2;
\end{equation}
(iii) the sixtic model~(\ref{eq:v6}) is singular on the curve $\gamma$ given by
\begin{equation}
  \label{eq:curve}
 5 g_2 g_6 = 2 g_4^2,\quad 4 g_4^3 = -225 g_6^2;
\end{equation}
and (iv) the model is in $G_1$ for $\mathbf{g}$ in $\Gamma-\gamma$ .

We apply these results to study the deformations of~(\ref{bmpp}) from the initial point $\mathbf{g}=(90,-15,1)$ in $\Gamma$.
The homogeneity in $T$ of~(\ref{eq:ec}) implies that the curve $\mathbf{g}/T$ lies on
$\Gamma$ for all $T>0$, while from~(\ref{eq:curve}) it follows that $\mathbf{g}/T\in \gamma$ only at $T= 60$.
Then  $\mathbf{g}/T\in G_1$ for all  $T\neq 60$ while $\mathbf{g}/60$ represents the multicritical
string model of~\cite{BR90} with potential function
\begin{equation}
  \label{bmp}
  V_c(\lambda) = \frac{3}{2} \lambda - \frac{1}{4} \lambda^2 + \frac{1}{60} \lambda^3.
\end{equation}
Finally, if we apply first the Bleher-Its deformation to $V(\lambda,\mathbf{g}/T)$ the resulting coupling parameters are
\begin{equation}
  g_2(T,t) = \left(1-\frac{1}{t}\right)+\frac{g_2}{t T},
  \qquad
  g_4(T,t) = \frac{g_4}{t^2 T},
  \qquad
  g_6(T,t) = \frac{g_6}{t^3 T}.
\end{equation}
For the particular values $g_2=90$, $g_4=-15$, $g_6=1$ of~(\ref{bmpp}) a direct computation shows that
\begin{equation}
  \frac{5}{2}  \frac{g_2(T,t) g_6(T,t)}{g_4(T,t)^2} = 1 + \frac{T(t-1)}{90},
\end{equation}
and using~(\ref{eq:in}) we find that $\mathbf{g}(T,t)\in G_1$ for all $T\neq 60$ and $t\geq 1$.
\section{Asymptotics of the recurrence coefficients\label{sec:asy}}
The main equation to determine the asymptotics of the recurrence coefficients is the
\emph{discrete string equation}~\cite{BL08}
\begin{equation}
  \label{str01}
  V_z(L)_{n,n-1}=\frac{n}{N}.
\end{equation}
Here $V_z$ stands for the derivative of the potential with respect to $z$
\begin{equation}
  V_z(z)=\sum_{k=1}^{2p} 2k g_{2k} z^{2k-1},
\end{equation}
the subindex $(n,n-1)$ denotes the corresponding matrix element between the orthogonal polynomials
\begin{equation}
   v_n(x)=P_{n,N}(x),\quad n\geq 0
\end{equation}
defined in~(\ref{pol}), and the operator $L$ acts on this family of polynomials as
\begin{equation}\label{lax}
L v_n=v_{n+1} + r_{n,N} v_{n-1},\quad r_{0,N}=0.
\end{equation}
The string equation~(\ref{str01}) can be written in the form
 \begin{equation}\label{stri2}
\oint_{\gamma}\frac{\rmd \lambda}{2\pi\rmi} V_{\lambda}(\lambda) U_{n, N}(\lambda)=\frac{n}{N}
\qquad(\lambda=z^2),
\end{equation}
where $\gamma$ is a large positively oriented circle $|\lambda|=R$, and $U_{n,N}$ is
the generating function
\begin{equation}\label{gen}
U_{n,N}(\lambda)=1+2 \sum_{k\geq 1}\left(L^{2k-1}\right)_{n,n-1} \lambda^{-k}.
\end{equation}
Note that $L$ is the Lax operator of the Toda hierarchy~\cite{GE91} and that $U_{n,N}$ is related to the resolvent
$\mathcal{R}(z)=(L-z)^{-1}$ of $L$ by
\begin{equation}
U_{n,N}(\lambda)=1-\mathcal{R}(z)_{n,n-1}-\mathcal{R}(-z)_{n,n-1}.
\end{equation}
In appendix~B we show that $U_{n,N}$  satisfies  the quadratic equation
\begin{equation}\label{res}
r_{n,N} \left(U_{n,N}+U_{n-1,N}\right)\left(U_{n,N}+U_{n+1,N}\right)=\lambda \left(U_{n,N}^2-1\right).
\end{equation}
\subsection{The continuum limit of the recurrence coefficient in the one-cut case}
We recall that our final goal is to solve the string equation~(\ref{stri2}) for the recurrence coefficient in the large $N$ limit.
If $\mathbf{g}\in G_1$ it has been rigorously established~\cite{BL05,DE99,KU00} that the asymptotics of the recurrence
coefficients as
\begin{equation}
  n\rightarrow \infty,
  \quad
  N\rightarrow \infty
  \quad
  \mbox{and}
  \quad
  \frac{n}{N}\rightarrow T
\end{equation}
in a neighborhood of $T=1$ is given by a series
\begin{equation}
  r_{n,N}(\mathbf{g}) \sim r(\epsilon,T,\mathbf{g})\quad (\epsilon=1/N),
\end{equation}
of the form
\begin{equation}
  \label{retg}
  r(\epsilon,T,\mathbf{g}) = \sum_{k\geq 0} r_k(T,\mathbf{g}) \epsilon^{2k}.
\end{equation}
In particular the leading coefficient is
\begin{equation}
  \label{r0}
  r_0 = \frac{\alpha^2}{4},
\end{equation}
where $(-\alpha,\alpha)$ is the eigenvalue support for the model $V(\lambda,\mathbf{g})/T$.
We write the  asymptotics of the generating function $U_{n,N}$ as a similar series
\begin{equation}
  \label{genc}
  U_{n,N}(\lambda) \sim U(\lambda,\epsilon;r),
\end{equation}
\begin{equation}
  \label{eq:ur}
  U(\lambda,\epsilon;r) = \sum_{k\geq 0} U_k(\lambda;r_0,\ldots,r_k) \epsilon^{2k}.
\end{equation}
Substituting the series~(\ref{retg}) and~(\ref{eq:ur}), and the corresponding shifted expansions
\begin{equation}
   r_{n+j,N}\sim r_{[j]}(\epsilon,T) =  r(\epsilon,T+j\epsilon),\quad k\in \mathbf{ Z},
\end{equation}
\begin{equation}
  U_{n+j,N}(\lambda) \sim U_{[j]}(\lambda,\epsilon;r) = U(\lambda,\epsilon;r_{[j]}),\quad k\in \mathbf{ Z},
\end{equation}
into~(\ref{res}), we get
\begin{equation}
  \label{resc}
  r \left(U+U_{[-1]}\right)\left(U+U_{[1]}\right) = \lambda\left(U^2-1\right).
\end{equation}
Incidentally, we note as a useful consequence of~(\ref{resc}) the linear equation
\begin{equation}\label{lins}
r_{[1]} \left(U_{[2]}+U_{[1]}\right) - r \left(U+U_{[-1]}\right) = \lambda \left(U_{[1]}-U\right).
\end{equation}

Identifying powers of $\epsilon$ recursively in~(\ref{resc}) or in~(\ref{lins}) we find that the coefficients $U_k$ can
be written in the form
\begin{equation}
  \label{str}
  U_k = U_0 \sum_{j=1}^{3k} \frac{ U_{k,j}(r_0,\ldots,r_k) }{ (\lambda-4r_0)^j },\quad  k\geq 1,
\end{equation}
where
\begin{equation}
  \label{eq:u0}
  U_0=\sqrt{\frac{\lambda}{\lambda-4 r_0}},
\end{equation}
and the functions $U_{k,j}(r_0,\ldots,r_k)$ are polynomials of degree $j$ in $r_0,\ldots, r_k$ and their
$T$ derivatives. Moreover, these polynomials are homogeneous of degree $2k$ with respect to the weight
$w(\partial_T^i r_j)=i+2j$, and the dependence of $U_k$ in $r_k$ comes solely from
\begin{equation}
  \label{eq:uk1}
  U_{k,1}=2 r_k,
\end{equation}
so that
\begin{equation}
  \label{rn}
  U_k = U_0 \left(\frac{2r_k}{\lambda-4r_0} + \cdots\right),
\end{equation}
where the dots stand for terms in $r_j,$ and their $T$ derivatives $r'_j,r''_j,\ldots$ with $ j=0,\ldots,k-1$. We give
explicitly the polynomials $U_{k,j}$ corresponding to $k=1$:
\begin{eqnarray}
  \nonumber U_{1,1} & = & 2 r_1,\\
                    U_{1,2} & = & 2 r_0  r_0'',\\
  \nonumber U_{1,3} & = & 10 r_0 (r_0')^2,
\end{eqnarray}
and to $k=2$:
\begin{eqnarray}
  \nonumber U_{2,1} & = & 2 r_2,\\
  \nonumber U_{2,2} & = & 2 r_1 r_0''+2 r_0r_1''+6 r_1^2+\frac{1}{6} r_0 r_0^{(4)},\\
  \nonumber U_{2,3} & = & 20 r_0 r_0' r_1' + \frac{22}{3}
                                           r_0 r_0' r_0^{(3)} + 10 r_1 (r_0')^2 + \frac{11}{2} r_0 (r_0'')^2\\
                                 &    &  {}+  20 r_1 r_0 r_0'' + 2  r_0^2 r_0^{(4)},\\
  \nonumber U_{2,4} & = & 140 r_0 (r_0')^2 r_0'' + 56  r_0{}^2 r_0' r_0^{(3)}+140 r_1 r_0 (r_0')^2 + 42 r_0^2 (r_0'')^2,\\
  \nonumber U_{2,5} & = & 924 r_0^2 (r_0')^2 r_0'' +378 r_0 (r_0')^4,\\
  \nonumber U_{2,6} & = & 2310 r_0^2 (r_0')^4.
\end{eqnarray}

Likewise, the continuum limit of the string equation~(\ref{stri2}) can be written as
\begin{equation}
  \label{stri22}
  \oint_{\gamma}\frac{\rmd \lambda}{2\pi\rmi} V_{\lambda}(\lambda) U(\lambda,\epsilon;r) = T
\end{equation}
or in terms of the expansion coefficients $U_k$,
\begin{equation}
  \label{stri3c}
  \oint_{\gamma}\frac{\rmd \lambda}{2\pi\rmi} V_{\lambda}(\lambda) U_k(\lambda;r_0,\ldots,r_k) =\delta_{k,0} T,\quad k\geq 0.
\end{equation}
Let us introduce the function
\begin{equation}
  W(r_0,\mathbf{g})
  =
  \oint_{\gamma}\frac{\rmd \lambda}{2\pi\rmi} V_{\lambda}(\lambda)
                                                                        \sqrt{\frac{\lambda}{\lambda-4 r_0}}
  = \sum_{n=1}^p {2 n \choose n}n g_{2n} r_0^n.
\end{equation}
Then, the $k=0$ equation in~(\ref{stri3c}) is
\begin{equation}
  \label{sin0}
  W(r_0,\mathbf{g})=T,
\end{equation}
or explicitly,
\begin{equation}
  \label{eq:eh}
  \sum_{n=1}^p {2 n \choose n}n g_{2n} r_0^n = T.
\end{equation}
This is an hodograph-type equation for $r_0$ (i.e., an equation which is linear in the independent variables
$g_2,\ldots,g_{2p},T$). Our next aim is to prove that~(\ref{stri3c}) permits the recursive calculation of the $r_k$
as functions of $T$ and $\mathbf{g}$.

For an even potential in the one-cut case, using the variable $\lambda=z^2$ and the relation between $\alpha$ and
$r_0$ given in equation~(\ref{r0}), we find that $w_1(z)=\sqrt{z^2-\alpha^2}=\sqrt{\lambda-4r_0}$. Therefore the
definition~(\ref{0.3}) of the polynomial $h(\lambda)$ can be written as
\begin{equation}
  2 \frac{\sqrt{\lambda} V_{\lambda}(\lambda)}{\sqrt{\lambda- 4 r_0}}
  = h(\lambda)+\mathcal{O}(\lambda^{-1}),\quad \lambda\rightarrow \infty.
\end{equation}
Consequently
\begin{equation}
  \label{cri1}
  \frac{\partial}{\partial r_0} W(r_0,\mathbf{g})
  = 2 \oint_{\gamma}\frac{\rmd \lambda}{2\pi\rmi}
        \frac{\sqrt{\lambda} V_{\lambda}}{(\lambda-4 r_0)^{3/2}}
  = h(\lambda)\Big|_{\lambda=4r_0}.
\end{equation}
Thus, given $\mathbf{g}_0\in G_1$ equation~(\ref{cri1}) implies that  $\partial_{r_0}W(r_0,\mathbf{g}_0)\neq 0$, and 
the hodograph equation~(\ref{sin0})  defines implicitly $r_0$ as a  locally smooth function  of $T$ and $\mathbf{g}$
in a neighborhood of $T_0=1$ and of $\mathbf{g}_0$. The remaining equations~(\ref{stri3c}) can be written as
\begin{equation}
  \label{ache}
  \sum_{j=1}^{3 k} W_j(r_0,\mathbf{g}) U_{k,j}(r_0,\ldots,r_k) = 0,\quad k\geq 1,
 \end{equation}
where
\begin{equation}
  \label{hj}
  W_j(r_0,\mathbf{g})
  =
  \oint_{\gamma}\frac{\rmd \lambda}{2\pi\rmi}
  \frac{\sqrt{\lambda} V_{\lambda}(\lambda)}{(\lambda-4 r_0)^{j+\frac{1}{2}}}
  =
  \frac{\partial_{r_0}^j W(r_0,\mathbf{g})}{2^j (2j-1)!!}.
\end{equation}
From~(\ref{eq:uk1}) and~(\ref{hj}) it follows that the term
$(\partial_{r_0}W(r_0,\mathbf{g})) r_k$ in equations~(\ref{ache}) equals a sum of terms of in $r_0,\ldots,r_{k-1}$
and their $T$ derivatives, and therefore define recursively  the coefficients $r_k$ as locally smooth functions
of $T$ and $\mathbf{g}$ in a neighborhood of $T_0=1$ and of $\mathbf{g}_0$.

We recall again that the coefficients $U_{k,j}(r_0,\ldots,r_k)$ in~(\ref{ache}) are polynomials of degree $j$ in
$r_0,\ldots,r_k$ and their $T$ derivatives. Repeated differentiation of the hodograph equation~(\ref{sin0}) with respect
to $T$ give the $T$ derivatives of $r_0$ as a rational function of $r_0$:
\begin{equation}
  r_0' = \frac{1}{\partial_{r_0} W},
  \qquad
  r_0'' = - \frac{\partial_{r_0}^2 W}{(\partial_{r_0} W)^3},
  \qquad
  \ldots
\end{equation}
and we can effectively solve equations~(\ref{ache}) for $r_k$ as a rational function of $r_0$.
Using the standard notation $W', W'',\ldots,W^{(j)}$ for the derivatives of $W$ with respect to $r_0$ we find
\begin{equation}
  \label{ere1}
  r_1 = r_0 \frac{2(W'')^2 - W' W'''}{12(W')^4},
\end{equation}
and
\begin{equation}
r_2 = r_0 \frac{X + r_0 Y}{1440 (W')^9},
\end{equation}
where
\begin{eqnarray}
 X & = & 700 W' \left(W''\right)^4-910 \left(W'\right)^2 \left(W''\right)^2 W'''+118 \left(W'\right)^3 \left(W'''\right)^2\nonumber\\
     &    & {}+180 \left(W'\right)^3 W'' W^{(4)}-18 \left(W'\right)^4 W^{(5)},
\end{eqnarray}
\begin{eqnarray}
Y & = & -980 \left(W''\right)^5+1760 W' \left(W''\right)^3 W'''-545 \left(W'\right)^2 W'' \left(W'''\right)^2\nonumber\\
   &     & {}-420 \left(W'\right)^2 \left(W''\right)^2 W^{(4)}+102 \left(W'\right)^3 W''' W^{(4)}\nonumber\\
   &     & {}+64 \left(W'\right)^3 W'' W^{(5)}-5 \left(W'\right)^4 W^{(6)},
\end{eqnarray}
which in turn show that the $r_k$ are rational functions of $r_0$. These expressions for $r_1$ and $r_2$
agree with Eq.~(4.25) and Eq.~(4.56) obtained by a different method in~\cite{SH95}. 

Finally, we remark that our method to calculate the coefficients $r_k$ of the large $N$ expansion ultimately depends
only on equations~(\ref{resc}) and~(\ref{stri22}), which are invariant under the symmetry transformation
\begin{equation}
  (\tilde{\epsilon}, \tilde{T}, \tilde{\mathbf{g}}) = \frac{1}{c}(\epsilon, T, \mathbf{g}), \quad c\geq 0.
\end{equation}
Hence it follows that $r(\tilde{\epsilon},\tilde{T},\tilde{\mathbf{g}})=r(\epsilon,T,\mathbf{g})$ and consequently
\begin{equation}
  \label{dept}
  r_k(T,\mathbf{g}) = \frac{1}{T^{2 k}} r_k(1, \mathbf{g}/T),\quad k\geq 0.
\end{equation}
\subsection{The recurrence coefficient under the  Bleher-Its deformation in the regular one-cut case}
Let $\mathbf{g}\in G_1$ such that its Bleher-Its deformation is in the regular one-cut case, i.e.,
$\mathbf{g}(t)\in G_1$ for all $t\geq 1$. Then we can apply the results of the previous subsection with $\mathbf{g}$
replaced by ${\mathbf{g}(t)}$ to conclude that in the limit~(\ref{lim})
\begin{equation}
  \label{et}
  r_{n,N}({\mathbf{g}(t)}) \sim r(\epsilon,T,{\mathbf{g}(t)})
\end{equation}
where  the coefficients $r_k(T,{\mathbf{g}(t)})$ of  the asymptotic series
\begin{equation}
  \label{one}
  r(\epsilon,T,{\mathbf{g}(t)}) = \sum_{k\geq 0} r_k(T,{\mathbf{g}(t)})\epsilon^{2k}
\end{equation}
are determined by
\begin{equation}
  \label{sin0a}
  W(r_0,{\mathbf{g}(t)}) = T,
\end{equation}
\begin{equation}
  \label{sin0b}
  \sum_{j=1}^{3 k} W_j(r_0,{\mathbf{g}(t)}) U_{k,j}(r_0,\ldots,r_k)=0,\quad k\geq 1,
\end{equation}
as smooth functions of $t$ and $T$  for $t\geq 1$ and $T$ near $T_0=1$. Our aim in this subsection
is to find a reformulation of~(\ref{sin0a})--(\ref{sin0b}) that decouples the dependence on $t$ and $\mathbf{g}$.

The string equation~(\ref{stri22}) for the deformed model is
\begin{equation}
  \oint_{\gamma}\frac{\rmd \lambda}{2\pi\rmi}
  V_{\lambda}(\lambda,{\mathbf{g}(t)}) U(\lambda,\epsilon;r) = T.
\end{equation}
If we substitute in this equation
\begin{equation}
  V_{\lambda}(\lambda,{\mathbf{g}(t)})
  =
  1 - \frac{1}{t} + \frac{1}{t} V_{\lambda} \left(\frac{\lambda}{t},\mathbf{g}\right),
\end{equation}
and take into account that $U\sim 1+2r/\lambda$ as $\lambda\rightarrow\infty$, we find that
\begin{equation}
  2 (t-1) \frac{r}{t} + \oint_{\gamma}\frac{\rmd \lambda}{2\pi\rmi}
                               \frac{1}{t} V_{\lambda}\left(\frac{\lambda}{t},\mathbf{g}\right) U(\lambda,\epsilon,r)
  = T,
\end{equation}
or with the change of variable $\lambda\to\lambda t$
\begin{equation}
  \label{stri22aa}
  2 (t-1) \frac{r}{t} + \oint_{\gamma}\frac{\rmd \lambda}{2\pi\rmi}
                               V_{\lambda}(\lambda,\mathbf{g}) U(\lambda t,\epsilon,r)
  = T.
\end{equation}
Note that the generating function $U(\lambda,\epsilon;r)$ is uniquely determined by~(\ref{resc}) and the asymptotic
behavior $U(\lambda,\epsilon;r)\sim 1$ as $\lambda\rightarrow\infty$. Since $U(\lambda t,\epsilon;r)$ satisfies~(\ref{resc})
with the substitution $r \rightarrow r/t$ and $U(\lambda t,\epsilon;r)\sim 1$ as $\lambda\rightarrow\infty$, we conclude that
$U(\lambda t,\epsilon;r)=U(\lambda,\epsilon; r/t)$. Alternatively, we can arrive at the same conclusion directly from the
explicit expressions~(\ref{str}) and~(\ref{eq:u0}) for the $U_k$ and from the fact that the functions $U_{k,j}$ are
polynomials of degree $j$ in $r_0,\ldots, r_k$ and their $T$ derivatives. Therefore, if we denote
\begin{equation}
  \label{lr}
  {\rt}(\epsilon,T,t,\mathbf{g})
  =
  \frac{r(\epsilon,T,{\mathbf{g}(t)})}{t}
  =
  \sum_{k\geq 0} {\rt}_k(T,t,\mathbf{g})\epsilon^{2k},
\end{equation}
equation~(\ref{stri22aa}) becomes
\begin{equation}
  \label{stri22aaa}
  2 (t-1) {\rt} + \oint_{\gamma}\frac{\rmd \lambda}{2\pi\rmi}
                      V_{\lambda}(\lambda,\mathbf{g}) U(\lambda,\epsilon,{\rt}) =T,
\end{equation}
or equivalently
\begin{equation}
  \label{rr0}
  2 (t-1) {\rt}_0 + W({\rt}_0,\mathbf{g}) =T,
\end{equation}
\begin{equation}
  \label{rrk}
  2(t-1) {\rt}_k + \sum_{j=1}^{3 k}W_j({\rt}_0,\mathbf{g}) U_{k,j}({\rt}_0,\ldots,{\rt}_k) =0,\quad k\geq 1.
\end{equation}
Note that the only changes introduced by the Bleher-Its deformation in our calculation of the recurrence coefficient
are the first term $2 (t-1) {\rt}_k$ and the substitution $r_i\rightarrow {\rt}_i$ in $U_{k,j}$.
\section{Topological expansions in the regular one-cut case\label{sec:top}}
In this section we implement our method to calculate the coefficients of the topological expansion
by means of our equations~(\ref{rr0})--(\ref{rrk}) and the Bleher-Its representation of the free energy,
which we repeat here for convenience:
\begin{equation}
  \label{eq:bioc}
  F_{N}(\mathbf{g})
  =
  F_N^{\mathrm{G}}
 +
  \int_1^{\infty} \frac{1-t}{t^2}
  \left[r_{N,N}({\mathbf{g}(t)}) \left( r_{N-1,N}({\mathbf{g}(t)})
                                                          + r_{N+1,N}({\mathbf{g}(t)})\right)-\frac{1}{2}\right]\rmd t.
\end{equation}
In the first subsection we derive integral expressions for the coefficients of the topological expansion.
Next, drawing on an idea that Bleher and Its~\cite{BL05} used to calculate the leading term $F^{(0)}(\mathbf{g})$
of the expansion, we give an efficient procedure to calculate higher-order coefficients and compute
explicit expressions of the first four coefficients for general models.
Finally, we apply our results to three widely studied models.
\subsection{Expressions for the coefficients of the topological expansion}
Equation~(\ref{eq:bioc}) involves only the recurrence coefficients $r_{N,N}$ and $r_{N\pm1,N}$. Therefore to study the
limit~(\ref{lim}) in the regular one-cut case we need only the asymptotic series for $r_{n,N}$ in a neighborhood of $T=1$:
\begin{equation}
  r_{N,N}({\mathbf{g}(t)}) \sim r(\epsilon,1,{\mathbf{g}(t)}) = t {\rt}(\epsilon,1,t,\mathbf{g}),
\end{equation}
\begin{equation}
  r_{N\pm1,N} \sim r(\epsilon, 1\pm\epsilon,{\mathbf{g}(t)})
  = t {\rt}(\epsilon,1\pm\epsilon,t,\mathbf{g}).
\end{equation}
Substituting these expansions in the Bleher-Its formula we have
\begin{equation}
  \label{bif1}
  F_{N}(\mathbf{g}) - F_N^{\mathrm{G}}
  \sim \int_1^{\infty} (1-t) f(\epsilon,t,\mathbf{g}) \rmd t,
\end{equation}
where (using again our shifting notation ${\rt}_{[\pm 1]}(\epsilon,1,t,\mathbf{g})={\rt}(\epsilon,1\pm\epsilon,t,\mathbf{g})$)
\begin{equation}
  \label{teta}
  f(\epsilon,t,\mathbf{g})
  = {\rt}(\epsilon,1,t,\mathbf{g})\left({\rt}_{[-1]}(\epsilon,1,t,\mathbf{g})+{\rt}_{[1]}(\epsilon,1,t,\mathbf{g})\right)-\frac{1}{2 t^2}.
\end{equation}
With the method discussed in the previous section we can readily obtain an expansion
\begin{equation}
  f(\epsilon,t,\mathbf{g}) = \sum_{k\geq 0} f_k(t,\mathbf{g}) \epsilon^{2k},
\end{equation}
where the first five coefficients are
\begin{eqnarray}
  \label{fs}
  f_0 & = & 2 {\rt}_0^2 - \frac{1}{2 t^2},\\
  \label{fs1}
  f_1 & = & {\rt}_0 (4 {\rt}_1+ {\rt}_0''),\\
  \label{fs2}
  f_2 & = & 2 {\rt}_1^2 + {\rt}_1 {\rt}_0'' + \frac{1}{12} {\rt}_0 (48 {\rt}_2+12 {\rt}_1'' + {\rt}_0''''),\\
 \label{fs3}
  f_3 & = & \rt_2 \rt_0''+\rt_1\,(4 \rt_2+\rt_1''+\frac{1}{12}\rt_0^{(4)})\nonumber \\
        &    &{}+\rt_0 (4\rt_3+\rt_2''+\frac{1}{360}  (30\rt_1^{(4)}+\rt_0^{(6)}),\\
  \label{fs4}
  f_4 &  = & 2 \rt_0\rt_4 + \rt_3 (2\rt_1+\rt_0'') + \rt_2 \left(2\rt_2+\rt_1''+\frac{1}{12}\rt_0^{(4)}\right)\nonumber\\
        &     &{}+ \rt_1\left(2 \rt_3+\rt_2''+\frac{30\rt_1^{(4)}+\rt_0^{(6)}}{360}\right) \nonumber\\
        &     &{}+\rt_0\left(2\rt_4+\rt_3''+\frac{1680 \rt_2^{(4)}+56\rt_1^{(6)}+\rt_0^{(8)}}{20160}\right).
\end{eqnarray}
Here the primes denote derivatives with respect to $T$ evaluated at $T=1$.
Note also that the term $f_k$ has weight $w(f_k)=2 k$. The  analysis of Bleher and Its in~\cite{BL05} shows that in the
regular one-cut case it is legitimate to perform term by term integration in~(\ref{bif1}), which yields the following
topological expansion of the free energy:
\begin{equation}
  \label{bif2}
  F_{N}(\mathbf{g}) - F_N^{\mathrm{G}} \sim \sum_{k\geq 0} F^{(k)}(\mathbf{g}) \epsilon^{2k},
\end{equation}
where
 \begin{equation}
   \label{bif2c}
   F^{(k)}(\mathbf{g}) = \int_1^{\infty} (1-t) f_k(t,\mathbf{g}) \rmd t.
\end{equation}
Therefore the direct method to calculate the coefficients of the topological expansion~(\ref{bif2}) is as follows:
first, use equations~(\ref{rr0})--(\ref{rrk}) to determine the coefficients ${\rt}_k$, then use equation~(\ref{teta})
to find the $f_k$, and finally perform the integration with respect to $t$ in equation~(\ref{bif2c}).
\subsection{Efficient calculation of the coefficients of the topological expansion}
The direct method to determine the coefficients of the topological expansion outlined in the preceding paragraph requires
explicit calculation of the functions ${\rt}_k(T,t,\mathbf{g})$ which is, except in the simplest cases, a difficult task.
In~\cite{BL05}  Bleher and Its used an ingenious idea to determine the leading coefficient $F^{(0)}(\mathbf{g})$
for general models $V(\lambda,\mathbf{g})$. In this section we will show that the same idea can be applied to evaluate
higher order coefficients $F^{(k)}(\mathbf{g})$.

It follows from our previous results that the functions $f_k(t,\mathbf{g})$ can be written as rational functions of
$t$ and ${\rt}_0(t,\mathbf{g})$. Now, if we denote
\begin{equation}
  \label{trans}
  \xi = {\rt}_0(1,t,\mathbf{g}),
\end{equation}
the hodograph equation~(\ref{rr0}) at $T=1$ implies that
\begin{equation}
  \label{trans1}
  t = 1 + \frac{1}{2\xi} (1-W(\xi,\mathbf{g})),
\end{equation}
which suggests to use $\xi$ as integration variable in~(\ref{bif2c}). At the lower limit of integration $t=1$ we have that
$\xi=r_0(1,\mathbf{g})$, while at the upper limit the new variable $\xi\sim 1/(2 t)\rightarrow 0$
as $t\rightarrow \infty$~\cite{BL05}. Therefore, with a trivial sign change absorbed in the definition of $R_k$,
equation~(\ref{bif2c}) can be written as
\begin{equation}
  \label{rat}
  F^{(k)}(\mathbf{g}) = \int_0^{r_0} R_k(\xi,\mathbf{g}) \rmd\xi,\quad k\geq 1,
\end{equation}
where the $R_k(\xi,\mathbf{g})$ are rational functions of $\xi$. Note that the evaluation of these integrals yields
the coefficients $F^{(k)}$ as functions of $\mathbf{g}$ and $r_0=r_0(1,\mathbf{g})$.

The presence of the term $1/(2 t^2)$ in the expression of $f_0$ requires a slightly different
integration process~\cite{BL05} to calculate $F^{(0)}(\mathbf{g})$. Using equations~(\ref{fs}), (\ref{trans})
and~(\ref{trans1}) we find:
\begin{eqnarray}
  F^{(0)}(\mathbf{g}) & = & \lim_{{\tau}\rightarrow\infty}
                                               \left[ 2 \int_{r_0}^{1/2{\tau}}(1-t(\xi)) \xi^2 \rmd t(\xi)
                                                       + \frac{1}{2} \ln{\tau} - \frac{1}{2}\right] \nonumber\\
                                   & = & \lim_{{\tau}\rightarrow\infty}
                                              \left[ \frac{1}{2} \ln \xi \Big|_{r_0}^{1/2{\tau}}
                                                     +
                                                     \frac{1}{2} \ln {\tau}
                                             \right]-\frac{1}{2}\nonumber\\
                                   &   & {}+  \int_0^{r_0} \left(W(\xi)-\frac{1}{2}W(\xi)^2\right)
                                                   \frac{\rmd\xi}{\xi}\nonumber\\
                                   &   & {} + \frac{1}{2} \int_0^{r_0} \left(W(\xi)-1\right) W'(\xi)\rmd\xi.
\end{eqnarray}
Hence, taking into account that $W(r_0,\mathbf{g})=1$ at $T=1$, it follows that
\begin{equation}
  \label{f0}
  F^{(0)}(\mathbf{g}) = - \frac{\ln r_0}{2}
                                       - \frac{\ln 2}{2}
                                       - \frac{3}{4}
                                       +\int_0^{r_0} \left(W-\frac{1}{2} W^2\right)\frac{\rmd\xi}{\xi}.
\end{equation}
Note that (since $W$ does not have a constant term) the integrand in the last term of~(\ref{f0})
is a polynomial in $\xi$.

Let us proceed now with the calculation of $F^{(1)}(\mathbf{g})$. We repeat here for convenience
equation~(\ref{rr0}) and particularize equation~(\ref{rrk}) for $k=1$:
\begin{equation}
  \label{rr0a}
  2 (t-1) {\rt}_0 + W({\rt}_0,\mathbf{g}) = T,
\end{equation}
\begin{equation}
  \label{rrkb}
  2 (t-1){\rt}_1 + \sum_{j=1}^{3}W_j({\rt}_0,\mathbf{g}) U_{1,j}({\rt}_0,{\rt}_1)=0.
\end{equation}
Differentiating with respect to $T$ in~(\ref{rr0a}) it follows that
\begin{equation}
  \rt_0'  =  \frac{1}{2(t-1+W_1)},
  \qquad
  \rt_0'' = -\frac{3W_2}{2(t-1+W_1)^3},
\end{equation}
and substituting these expressions in~(\ref{rrkb}) we find
\begin{equation}
  \rt_1 = \rt_0 \left[ \frac{3W_2^2}{2(t-1+W_1)^4} - 
                               \frac{5W_3}{4(t-1+W_1)^3}
                       \right].
\end{equation}
Therefore
\begin{equation}
  f_1 = \rt_0 (4\rt_1+\rt_0'')
        = \rt_0 \left[ \frac{6 W_2^2 \rt_0}{(t-1+W_1)^4} -
                             \frac{10W_3 \rt_0 + 3 W_2}{2 (t-1+W_1)^3}
                    \right],
\end{equation}
and changing the variable from $t$ to $\xi$ with~(\ref{trans1}) we obtain the following expression for $F^{(1)}(\mathbf{g})$:
\begin{equation}
  \label{f1n}
  F^{(1)}(\mathbf{g})
  =
  \int_0^{r_0} (W-1) \left[ \frac{\xi^2 24 W_2^2}{(1+2\xi W_1-W)^3} -
                                           \frac{10\xi W_3+3 W_2}{(1+2\xi W_1-W)^2}\right]\xi\rmd\xi.
\end{equation}
Using~(\ref{hj}) we can identify the integrand as a total derivative:
\begin{eqnarray}
  \frac{\xi  (-1+W) W'' \left(-3+3 W-3 \xi  W'+2 \xi ^2 W''\right)}{12 \left(1-W+\xi  W'\right)^3}
  -
  \frac{\xi ^2 (-1+W) W'''}{12 \left(1-W+\xi  W'\right)^2}\nonumber\\
 = 
 \frac{1}{12}\frac{\rmd}{\rmd\xi}
 \left(\ln \left(-\xi  W'+W-1\right)-\frac{(W-1) \xi ^2 W''}{\left(-\xi  W'+W-1\right)^2}\right).
\end{eqnarray}
Thus we arrive at the result
\begin{equation}
  \label{sf1}
  F^{(1)}(\mathbf{g}) =  \frac{1}{12}\ln\left(r_0 W'(r_0,\mathbf{g})\right),
\end{equation}
in agreement with Eq.~(4.27) of~\cite{SH96}.

The calculation of the next coefficient $F^{(2)}(\mathbf{g})$ of the topological expansion is entirely similar, and we omit
the intermediate steps which are easily performed with a symbolic computation program.
Using~(\ref{rr0}),~(\ref{rrk}) for $k=1,2$ and~(\ref{fs2}) we find
\begin{eqnarray}
R_2(\xi,\mathbf{g})  &= &  3(1-W)
                                         \left[\frac{\xi^9}{\sigma^8} 56448 W_2^5\right.\nonumber\\
   & &{}- \frac{\xi^7}{\sigma^7} \left(10368 W_2^4 + 84480 \xi  W_2^3 W_3\right)\nonumber\\
   & &{}+  \frac{\xi^5}{\sigma^6} \left[420 W_2^3+10800 \xi  W_2^2 W_3 \right.\nonumber\\
   & &\quad\left. {}+ \xi ^2 \left(21800 W_2 W_3^2+23520 W_2^2 W_4\right)\right]\nonumber\\
   & &{}-  \frac{\xi^4}{\sigma^5} \left[260 W_2 W_3+\xi  \left(1110 W_3^2+2380 W_2 W_4\right)\right.\nonumber\\
   & &\quad\left. {}+ \xi ^2 \left(4760 W_3 W_4+5376 W_2 W_5\right)\right]\nonumber\\
   & &\left.{}+  \frac{\xi^3}{\sigma^4} \left(35 W_4+336 \xi  W_5+770 \xi ^2 W_6\right)\right].
\end{eqnarray}
where
\begin{equation}
  \sigma = 1-W+2 \xi  W_1 = 1-W +\xi  W'.
\end{equation}
Using~(\ref{hj}) we can again identify $ R_2(\xi,\mathbf{g})$ as a total derivative,
\begin{eqnarray}
  R_2(\xi,\mathbf{g}) & = & -\frac{1}{2880} \frac{\rmd}{\rmd\xi}\left[
                 -\frac{\xi^8}{\sigma^7} 280(-1+W) (W'')^4\right.\nonumber\\
     &     & {}+\frac{\xi^6}{\sigma^6} (-1+W) \left(300 (W'')^3+400 \xi (W'')^2 W'''\right)\nonumber\\
     &     & {}-\frac{\xi^5}{\sigma^5} \left[56 \xi (W'')^3+(-1+W)\times\right.\nonumber\\
     &     &\qquad\left.\left(260 W'' W'''+58 \xi (W''')^2+88 \xi W'' W^{(4)}\right)\right]\nonumber\\
     &     & {}+\frac{\xi^4}{\sigma^4}\left[-9 (W'')^2+58 \xi W'' W'''\right.\nonumber\\
     &     &\quad\left.{}+(-1+W) \left(36 W^{(4)}+10 \xi W^{(5)}\right)\right]\nonumber\\
     &     &\left. {}+\frac{\xi^2}{\sigma^3}\left(-12  W''+4 \xi W'''-10 \xi^2 W^{(4)}\right)-\frac{12}{\sigma^2}\right],
\end{eqnarray}
and we finally obtain
\begin{eqnarray}\label{sf2}
  F^{(2)}(\mathbf{g})
               & = & -\frac{1}{240} + \frac{1}{240 r_0^2 W'(r_0)^2} + \frac{7 {r_0} W''(r_0)^3}{360 W'(r_0)^5}\nonumber\\
               &   & {}+\frac{W''(r_0) \left(9 W''(r_0)-58 {r_0} W^{(3)}(r_0)\right)}{2880 W'(r_0)^4}\nonumber\\
               &   & {}+\frac{6 W''(r_0)-2 {r_0} W^{(3)}(r_0)+5 {r_0}^2 W^{(4)}(r_0)}{1440 {r_0} W'(r_0)^3}.
\end{eqnarray}
This expression agrees with the result of~\cite{SH96} (up to a trivial mistake in the sign of the first fraction of
his~Eq.~(1.18), which is fixed in his application to the quartic model in~Eq.~(4.59)). 
Our method can be easily carried on further with a symbolic computation program. For example,
the next coefficient turns out to be
\begin{eqnarray}
  \label{sf3}
  F^{(3)}(\mathbf{g}) & = &
   \frac{1}{1008}
   -\frac{1}{1008 r_0^4W'(r_0)^4}-\frac{W''(r_0)}{504 r_0^3W'(r_0)^5}\nonumber\\
   &&{}-\frac{1}{6048 r_0^2W'(r_0)^6}
   \Big[15 W''(r_0)^2-4W^{(3)}(r_0) W'(r_0)\Big]\nonumber\\
   &&{}-\frac{1}{6048 r_0 W'(r_0)^7}
   \Big[15W''(r_0)^3+W^{(4)}(r_0)W'(r_0)^2\nonumber\\
   &&\quad{}-10 W^{(3)}(r_0) W'(r_0)W''(r_0)\Big]\nonumber\\
   &&{}-\frac{1}{725760W'(r_0)^8}
   \Big[1575W''(r_0)^4-24 W^{(5)}(r_0)W'(r_0)^3\nonumber\\
   &&\quad{}+200 W^{(3)}(r_0)^2W'(r_0)^2+300 W^{(4)}(r_0)W'(r_0)^2 W''(r_0)\nonumber\\
   &&\quad{}-1800 W^{(3)}(r_0)W'(r_0) W''(r_0)^2\Big]\nonumber\\
   &&{}-\frac{r_0}{362880W'(r_0)^9}
   \Big[-21420 W''(r_0)^5-133 W^{(6)}(r_0)W'(r_0)^4\nonumber\\
   &&\quad{}+1644 W^{(5)}(r_0)W'(r_0)^3 W''(r_0)\nonumber\\
   &&\quad{}+2488 W^{(3)}(r_0)W^{(4)}(r_0) W'(r_0)^3\nonumber\\
   &&\quad{}-10170W^{(4)}(r_0) W'(r_0)^2W''(r_0)^2\nonumber\\
   &&\quad{}+40110 W^{(3)}(r_0) W'(r_0)W''(r_0)^3\nonumber\\
   &&\quad{}-12783 W^{(3)}(r_0)^2
   W'(r_0)^2 W''(r_0)\Big]\nonumber\\
   &&{}-\frac{r_0^2}{362880W'(r_0)^{10}}
   \Big[34300 W''(r_0)^6-35 W^{(7)}(r_0)W'(r_0)^5\nonumber\\
   &&\quad{}+607 W^{(4)}(r_0)^2W'(r_0)^4-2915 W^{(3)}(r_0)^3W'(r_0)^3\nonumber\\
   &&\quad{}+539 W^{(6)}(r_0)W'(r_0)^4 W''(r_0)\nonumber\\
   &&\quad{}+1006 W^{(3)}(r_0)W^{(5)}(r_0) W'(r_0)^4\nonumber\\
   &&\quad{}-4284W^{(5)}(r_0) W'(r_0)^3W''(r_0)^2\nonumber\\
   &&\quad{}+22260 W^{(4)}(r_0)W'(r_0)^2 W''(r_0)^3\nonumber\\
   &&\quad{}-81060W^{(3)}(r_0) W'(r_0) W''(r_0)^4\nonumber\\
   &&\quad{}+43050W^{(3)}(r_0)^2 W'(r_0)^2W''(r_0)^2\nonumber\\
   &&\quad{}-13452 W^{(3)}(r_0)W^{(4)}(r_0) W'(r_0)^3W''(r_0)\Big].
\end{eqnarray}
This expression reduces to the result found by Shirokura except for the replacement of $388$
by $300$ in the fourth coefficient of $E_0^{(3)}$ in Eq.~(50) of~\cite{SH95}.

We apply now (\ref{f0}), (\ref{sf1}) and~(\ref{sf2}) to obtain explicit expressions for the first three coefficients
of the quartic, two-valence and sixtic models (we omit the lengthy expressions for $F^{(3)}(\mathbf{g})$
which are obtained in exactly the same way using~(\ref{sf3})).
\subsubsection{The quartic model in the regular one-cut case}
We have shown in section~\ref{sec:qm} that if $(g_2,g_4)\in G_1$ the quartic model~(\ref{bip}) is in the regular
one-cut case. Using equations~(\ref{f0}), (\ref{sf1}) and~(\ref{sf2}) with
\begin{equation}
  \label{eq:w4app}
  W(r_0,\mathbf{g}) = 2 g_2 r_0 + 12 g_4 r_0^2
\end{equation}
we find
\begin{equation}
  \label{eq:qf0}
  F^{(0)}(\mathbf{g}) = - \frac{3}{8} 
                                   + \frac{5}{6}\left(g_2 r_0\right)
                                   - \frac{1}{6}\left(g_2 r_0\right)^2
                                   - \frac{1}{2} \ln\left(2 r_0\right),
\end{equation}
\begin{equation}
  \label{eq:qf1}
  F^{(1)}(\mathbf{g}) = \frac{1}{12} \ln\left(2\left(1-g_2 r_0\right)\right),
\end{equation}
\begin{equation}
  \label{eq:qf2}
 F^{(2)}(\mathbf{g}) = \frac{(2 g_2 r_0-1)^3 \left(41 + 21 g_2 r_0-6 (g_2 r_0)^2\right)}{11520 (1-g_2 r_0)^5},
\end{equation}
where $r_0$ is the positive root of the hodograph equation~(\ref{eq:eh})
\begin{equation}
  2 g_2 r_0 + 12 g_4 r_0^2 = 1,
\end{equation}
namely
\begin{equation}
  r_0 = \frac{- g_2 + \sqrt{g_2^2+12 g_4}}{12 g_4}.
\end{equation}
\subsubsection{Two-valence models}
For the two-valence models
\begin{equation}
  \label{2v}
  V(\lambda,\mathbf{g}) = g_2 \lambda + g_{2\nu} \lambda^{\nu}, \quad \nu\geq 2,
\end{equation}
in the region $g_2>0$, $g_{2 \nu}>0$, the Bleher-Its deformed potential is a convex function of $z$
for all $t\geq 1$, and therefore these models are in the regular one-cut case. Using
\begin{equation}
  W(r_0,\mathbf{g}) = 2 g_2 r_0 + \nu{2\nu\choose\nu} g_{2\nu} r_0^{\nu}
\end{equation}
we find:
\begin{equation}
  \label{F0erc}
  F^{(0)}(\mathbf{g}) =  - \frac{3(\nu-1)}{4\nu}
                                    + \frac{(\nu-1)(2\nu+1)}{\nu(\nu+1)} (g_2 r_0)
                                    - \frac{(\nu-1)^2}{\nu(\nu+1)} (g_2 r_0)^2
                                    - \frac{1}{2} \ln(2 r_0),
\end{equation}
\begin{equation}
  \label{F12}
  F^{(1)}(\mathbf{g}) = \frac{1}{12}\ln\left(\nu-(\nu-1)2g_{2}r_0\right),
\end{equation}
\begin{eqnarray}
  \label{F22}
  F^{(2)}(\mathbf{g}) & = &  \frac{(2g_2r_0-1)(\nu-1)}{2880(\nu-2(\nu-1)g_2r_0)^5}
                                     \times\nonumber\\
                  &   & \Big[- \nu^3 (8 \nu ^2 + 5\nu -1) + 2 \nu^2(\nu-1) (16 \nu^2 + 40\nu -1) (g_2 r_0 )
                           \nonumber\\
                  &    & {}- 4\nu(\nu-1)^2 (8 \nu ^2-\nu+44) (g_2 r_0)^2
                           \nonumber\\
                  &    & {}- 96 (\nu-1)^3 (4 \nu +1)  (g_2 r_0)^3
                             + 192 (\nu-1)^4 (g_2 r_0)^4 \Big].                             
\end{eqnarray}
Here $r_0$ is determined as the positive root of the hodograph equation
\begin{equation}
  2 g_2 r_0 + \nu{2\nu\choose\nu} g_{2\nu} r_0^{\nu} = 1.
\end{equation}
Note that (\ref{F0erc})--(\ref{F22}) reduce to the quartic results for $\nu=2$. For $g_2=1/2$ the
expressions~(\ref{F0erc})--(\ref{F22}) reduce to the corresponding results in~\cite{ER08,ER09}.
\subsubsection{Sixtic models}
Our last application concerns sixtic potentials
\begin{equation}
  \label{sixtic}
  V(\lambda,\mathbf{g}) = g_2\lambda + g_4\lambda^2 + g_6\lambda^3
\end{equation}
in the region $g_2>0$, $g_4>0$, $g_6>0$. Again the convexity argument proves
that these models are in the regular one-cut case. Using
\begin{equation}
  W(r_0,\mathbf{g}) = 2 g_2 r_0 +  12 g_4 r_0^2+ 60 g_6 r_0^3 
\end{equation}
we find:
\begin{equation}
  \label{six0}
  F^{(0)}(\mathbf{g}) =  - \frac{1}{2}
                                  + \frac{7}{6} (g_2 r_0)
                                  - \frac{1}{3} (g_2 r_0)^2
                                  + \frac{8}{5}  (g_4 r_0^2)
                                  - \frac{6}{5}   (g_4 r_0^2)^2 
                                  - \frac{6}{5} g_2 g_4 r_0^3
                                  -\frac{1}{2} \ln \left(2 r_0\right),
\end{equation}
\begin{equation}
  \label{six1}
  F^{(1)}(\mathbf{g}) = \frac{1}{12} \ln\left(3-4 g_2 r_0-12 g_4 r_0^2\right),
\end{equation}
\begin{eqnarray}
  \label{six3}
  F^{(2)}(\mathbf{g}) & = & - \frac{1}{240}
                                          + \frac{593}{720 \left(12 g_4 r_0^2+4 g_2r_0-3\right)^2}\nonumber\\
                                 &   & {}+ \frac{169 g_2^2+2928 g_4-1716 g_4g_2r_0}{
                                                     720 g_4 \left(12 g_4 r_0^2+4 g_2 r_0-3\right)^3}\nonumber\\
                                &   &  {}+ \frac{224 g_2^4+7587 g_4 g_2^2+45765 g_4^2-(57888 g_4^2g_2+6756 g_4g_2^3)r_0}{
                                                       6480 g_4^2 \left(12 g_4 r_0^2+4 g_2 r_0-3\right)^4}\nonumber\\
                                &   & {}+\frac{7 \left(6 g_2^4+81 g_4g_2^2+243 g_4^2-(8g_2^5+126 g_4g_2^3+486 g_4^2g_2)r_0\right)}{
                                                     405 g_4^2 \left(12 g_4 r_0^2+4 g_2 r_0-3\right)^5}.\nonumber\\
                                & &
\end{eqnarray}
In this case $r_0$ is the positive root of the hodograph equation
\begin{equation}
   2 g_2 r_0 + 12 g_4 r_0^2 + 60 g_6 r_0^3 = 1.
\end{equation}
\section{Counting numbers}
The methods of Ercolani and McLaughlin~\cite{ER03} prove the existence of the topological expansion
for matrix models
\begin{equation}
  V_{EM}(\lambda) = \frac{\lambda}{2} + \sum_{j=1}^{\nu}t_{2j} \lambda^{j},
\end{equation}
under the hypothesis that there exists a path in the space of coupling constants connecting
$\mathbf{t}=(t_2,t_4,\ldots,t_{2\nu})$ to the origin $\mathbf{0}$. The corresponding coefficients
$F^{(k)}(\mathbf{t})$  are analytic functions of $\mathbf{t}$ near the origin and their Taylor
expansions determine graphical enumeration numbers.

To put these results in context, we briefly recall that a $k$-map is a graph which is embedded into a surface
of genus $k$ in such a way that (i) the edges do not intersect and (ii) dissecting the surface along the edges
decomposes it into faces which are homeomorphic to a disk. We can formulate the result
of~\cite{ER03} as the following representation: if
\begin{equation}
  \label{ercf}
  F^{(k)}(\mathbf{t}) = -\sum_{n_{2j}\geq 1}
                                    \frac{1}{n_2!\ldots n_{2p}!}
                                    (-t_2)^{n_2}\cdots(-t_{2p})^{n_{2p}} \kappa_k(n_2,\ldots,n_{2p}),
\end{equation}
then $\kappa_k(n_2,\dots,n_{2p})$ is the number of connected $k$-maps with a number $n_{2j}$
of $2j$-valent vertices in which all the vertices are labeled as distinct and all the edges emanating
from each vertex are labelled as distinct as well~\cite{DI06,ER03,KO09}.

It is straightforward to rephrase our results of the previous section in the notation of~(\ref{ercf}) and therefore
we have a direct method to calculate the $\kappa_k(n_2,\dots,n_{2p})$: we introduce the change of variable
$\lambda'=\lambda/2$ to reduce $V_{EM}(\lambda)$ to the form~(\ref{eve}), which in turn implies the following
relation between our set of $\mathbf{g}$ coupling constants and $\mathbf{t}$:
\begin{equation}
  \label{gt}
  g_{2k}(\mathbf{t}) = \delta_{k,2} + 2^k t_{2k},\quad k=1,\ldots,p.
\end{equation}
As an application of this first, direct procedure, we have carried out these substitutions in the
topological expansion of the quartic model~(\ref{eq:qf0})--(\ref{eq:qf2}), expanded in Taylor series
these coefficients, and found the corresponding counting numbers $\kappa_k(n_2,n_4)$, the
first of which we present in table~\ref{tab:qcn}.
\begin{table}
\caption{Lowest three counting numbers $\kappa_k(n_2,n_4)$ for the quartic model
             $V(\lambda)=g_2\lambda+g_4\lambda^2$ and all combinations of the $n_i$
             up to $4$.\label{tab:qcn}}
\begin{center}
\begin{tabular}{rrrrrr}
\hline\hline
\multicolumn{1}{c}{$n_2$} &
\multicolumn{1}{c}{$n_4$} &
\multicolumn{1}{c}{$\kappa_0(n_2,n_4)$} &
\multicolumn{1}{c}{$\kappa_1(n_2,n_4)$} &
\multicolumn{1}{c}{$\kappa_2(n_2,n_4)$}\\
\hline
 0 & 0 & 0  & 0  & 0\\
 0 & 1 &  2  &  1  & 0\\
 0 & 2 &  36  & 60  & 0\\
 0 & 3 &  1728  & 6336  & 1440\\
 0 & 4 &   145152 & 964224 & 770688\\
 \hline
 1 & 0 & 1  & 0  & 0\\
 1 & 1 &  8  & 4  & 0\\
 1 & 2 &  288  & 480  & 0\\
 1 & 3 & 20736   & 76032  & 17280\\
 1 & 4 &   2322432 &  15427584 & 12331008\\
 \hline
 2 & 0 & 2  & 0  & 0\\
 2 & 1 &  48  & 24  & 0\\
 2 & 2 &  2880  & 4800  & 0\\
 2 & 3 & 290304   & 1064448  & 241920\\
 2 & 4 & 41803776  & 277696512 & 221958144\\
 \hline
 3 & 0 & 8  & 0  & 0\\
 3 & 1 &  384  & 192  & 0\\
 3 & 2 &  34560  & 57600  & 0\\
 3 & 3 & 4644864   & 17031168  & 3870720\\
 3 & 4 & 836075520  & 5553930240 & 4439162880\\
 \hline
 4 & 0 & 48  &  0 & 0\\
 4 & 1 &  3840  & 1920  & 0\\
 4 & 2 & 483840  & 806400 & 0\\
 4 & 3 & 83607552  & 306561024  & 69672960\\
 4 & 4 &  18393661440 & 122186465280 & 97661583360\\
\hline\hline
\end{tabular}
\end{center}
\end{table}

However, since the counting numbers depend on the coefficients of the Taylor expansion of the $F^{(k)}$,
drawing on our analysis of the previous section we can calculate directly these Taylor coefficients without
requiring  the explicit evaluation  of the $F^{(k)}$ themselves, which turns out to be much more efficient.
In essence, the idea is to obtain first the Taylor expansion in $\mathbf{t}$ of the integrand of~(\ref{rat}),
and subsequently to perform the integration in~(\ref{bif2c}) with respect to the Bleher-Its deformation
parameter $t$ term by term. A similar approach has been recently used in~\cite{BL10} to investigate
counting numbers in the cubic model. In detail, the procedure is as follows:
\begin{enumerate}
\item Use ~(\ref{gt})  to write Eq.~(\ref{rr0}) at $T=1$ as
         \begin{equation}
         \label{rr0t}
           \rt_0+\sum_{j=1}^p\left(\begin{array}{c} 2j \\ j \end{array}\right)j2^{j-1}\frac{t_{2j}}{t}\rt_0^j=\frac{1}{2t},
         \end{equation}
         and then use implicit differentiation to calculate the Taylor expansion at $\mathbf{t}=\mathbf{0}$
         \begin{equation}
         \label{r0seriet}
            \rt_0 = \frac{1}{2t} 
           + \sum_{\begin{array}{c}j_1,\dots,j_p\geq0\\j_1+\cdots+j_p\geq1\end{array}}
           \frac{c_{j_1 j_2\dots j_p}}{t^{1+j_1+2j_2+\cdots+pj_p}}t_2^{j_1}t_4^{j2}\cdots t_{2p}^{j_p}.
         \end{equation}
\item Use the string equations~(\ref{rr0})--(\ref{rrk})  to write $f_k$ as rational functions of $\rt_0$.
         Then substitute~(\ref{r0seriet}) in the resulting expressions  and determine the Taylor expansion
         of $f_k$  at $\mathbf{t}=\mathbf{0}$. 
 \item Perform the integration
          \begin{equation}
            F^{(k)}(\mathbf{t})=\int_1^{\infty}(1-t)f_k(t,\mathbf{t})dt,
          \end{equation}
          term by term and find the numbers $\kappa_k(n_2,\ldots,n_{2p})$.
\end{enumerate}

We have implemented this strategy for the two-valence and for the sixtic models, and present some
of our results in table~\ref{tab:scn}. The $\kappa_k(n_2,\ldots,n_{2\nu})$ grow quickly in number
and in magnitude, and we give a complete table only up to $n_i=2$. Note that the results in
table~\ref{tab:scn} with $n_6=0$ agree with the corresponding results in table~\ref{tab:qcn}.
Our results also agree with those for $\kappa_1(n_2,0,\dots,n_{2\nu})$ and
$\kappa_2(n_2,0,\dots,n_{2\nu})$ in~\cite{ER09}. Similarly, in table~\ref{tab:scnk4}
we present all the fifth nonvanishing counting numbers $\kappa_4(n_2,n_4,n_6)$ with
$0\le n_i\le 3$ for the sixtic model.
\begin{table}
\caption{Lowest four counting numbers $\kappa_k(n_2,n_4,n_6)$ for the sixtic model
             $V(\lambda)=g_2\lambda+g_4\lambda^2+g_6\lambda^3$ and all combinations
             of the $n_i$ up to $2$.\label{tab:scn}}
\begin{center}
\begin{tabular}{rrrrrrr}
\hline\hline
\multicolumn{1}{c}{$n_2$} &
\multicolumn{1}{c}{$n_4$} &
\multicolumn{1}{c}{$n_6$} &
\multicolumn{1}{c}{$\kappa_0(n_2,n_4,n_6)$} &
\multicolumn{1}{c}{$\kappa_1(n_2,n_4,n_6)$} &
\multicolumn{1}{c}{$\kappa_2(n_2,n_4,n_6)$} &
\multicolumn{1}{c}{$\kappa_3(n_2,n_4,n_6)$}\\
\hline
 0 & 0 & 0 & 0 & 0 & 0 & 0\\
 0 & 0 & 1 & 5 & 10 & 0 & 0\\
 0 & 0 & 2 & 600 & 4800 & 4770 & 0\\
 0 & 1 & 0 & 2 & 1 & 0 & 0\\
 0 & 1 & 1 & 144 & 600 & 156 & 0\\
 0 & 1 & 2 & 43200 & 540000 & 1161360 & 224280\\
 0 & 2 & 0 & 36 & 60 & 0 & 0\\
 0 & 2 & 1 & 8640 & 63360 & 56160 & 0\\
 0 & 2 & 2 & 4665600 & 85190400 & 329002560 & 217339200\\
 \hline
 1 & 0 & 0 & 1 & 0 & 0 & 0\\
 1 & 0 & 1 & 30 & 60 & 0 & 0\\
 1 & 0 & 2 & 7200 & 57600 & 57240 & 0\\
 1 & 1 & 0 & 8 & 4 & 0 & 0\\
 1 & 1 & 1 & 1440 & 6000 & 1560 & 0\\
 1 & 1 & 2 & 691200 & 8640000 & 18581760 & 3588480\\
 1 & 2 & 0 & 288 & 480 & 0 & 0\\
 1 & 2 & 1 & 120960 & 887040 & 786240 & 0\\
 1 & 2 & 2 & 93312000 & 1703808000 & 6580051200 & 4346784000\\
 \hline
 2 & 0 & 0 & 2 & 0 & 0 & 0\\
 2 & 0 & 1 & 240 & 480 & 0 & 0\\
 2 & 0 & 2 & 100800 & 806400 & 801360 & 0\\
 2 & 1 & 0 & 48 & 24 & 0 & 0\\
 2 & 1 & 1 & 17280 & 72000 & 18720 & 0\\
 2 & 1 & 2 & 12441600 & 155520000 & 334471680 & 64592640\\
 2 & 2 & 0 & 2880 & 4800 & 0 & 0\\
 2 & 2 & 1 & 1935360 & 14192640 & 12579840 & 0\\
 2 & 2 & 2 & 2052864000 & 37483776000 & 144761126400 & 95629248000\\
\hline\hline
\end{tabular}
\end{center}
\end{table}
\begin{table}
\caption{Nonvanishing counting numbers $\kappa_4(n_2,n_4,n_6)$ with $0\le n_i\le 3$ for the sixtic model
             $V(\lambda)=g_2\lambda+g_4\lambda^2+g_6\lambda^3$.\label{tab:scnk4}}
\begin{center}
\begin{tabular}{rrrrrrr}
\hline\hline
\multicolumn{1}{c}{$n_2$} &
\multicolumn{1}{c}{$n_4$} &
\multicolumn{1}{c}{$n_6$} &
\multicolumn{1}{c}{$\kappa_4(n_2,n_4,n_6)$}\\
\hline
 0 & 1 & 3 & 1143525600 \\
 0 & 2 & 3 & 2201217638400 \\
 0 & 3 & 2 & 24069830400 \\
 0 & 3 & 3 & 2836746385920000 \\
\hline
 1 & 1 & 3 & 25157563200 \\
 1 & 2 & 3 & 57231658598400 \\
 1 & 3 & 2 & 577675929600 \\
 1 & 3 & 3 & 85102391577600000 \\
\hline
 2 & 1 & 3 & 603781516800 \\
 2 & 2 & 3 & 1602486440755200 \\
 2 & 3 & 2 & 15019574169600 \\
 2 & 3 & 3 & 2723276530483200000 \\
\hline
 3 & 1 & 3 & 15698319436800 \\
 3 & 2 & 3 & 48074593222656000 \\
 3 & 3 & 2 & 420548076748800 \\
 3 & 3 & 3 & 92591402036428800000 \\
 \hline\hline
\end{tabular}
\end{center}
\end{table}
\section{Singular one-cut cases}
Let $\mathbf{g}\in G$ be such that its Bleher-Its deformation is in the singular one-cut case, i.e.,
$\mathbf{g}(t)\in G_1$ for all $t>1$ but $\mathbf{g}=\mathbf{g}(1)$ determines a singular model.
Our discussion in sections~\ref{sec:asy} and~\ref{sec:top} shows that the integrals for the coefficient
of the topological expansion~(\ref{bif2})--(\ref{bif2c}) converge provided that the function
${\rt}_0(T,t,\mathbf{g})$ is smooth near $(T,t)=(1,1)$. For this type of singular cases the topological
expansion~(\ref{bif2})--(\ref{bif2c}) still exists. An example of this situation is the quartic model~(\ref{bip})
for $g_2 = -2\sqrt{g_4}$.

However, in the singular one-cut cases where the function ${\rt}_0(T,t,\mathbf{g})$ is not smooth
near $(T,t)=(1,1)$ the topological expansion~(\ref{bif2})--(\ref{bif2c}) is ill-defined, because the
integrals defining the coefficients $F^{(k)}$ for $k\geq 1$ diverge.
This is the case of the Brezin-Marinari-Parisi model~(\ref{bmp}). We next discuss this critical behavior
in general and present a method of regularization.
\subsection{Critical behavior and a triple-scaling method of regularization}
Let us consider a singular one-cut deformation such that the string equation~(\ref{stri22}) has
a critical point  of order $m\geq 2$ at $(r_0,T)=(r_c,1)$. That it to say,
\begin{equation}
  \label{mcr1}
  W(r_c,\mathbf{g}) =1,
\end{equation}
\begin{equation}
  \label{mcr2}
  \partial_{r_0} W(r_c,\mathbf{g}) = \cdots =\partial_{r_0}^{m-1}W(r_c,\mathbf{g}) = 0,
  \qquad
  \partial_{r_0}^m W(r_c,\mathbf{g}) \neq 0.
\end{equation}
In this case the implicit function theorem does not apply to equation~(\ref{rr0}) near
$(T_0,t_0)=(1,1)$ with ${\rt}_0(1,1,\mathbf{g})=r_c$. In fact, ${\rt}_0(T,1,\mathbf{g})$ can be
expanded in powers of $ (T-1)^{1/m}$ and, provided that $r_c\neq 0$, ${\rt}_0(1,t,\mathbf{g})$
can be also expanded in  powers of $ (t-1)^{1/m}$. As a consequence, the system~(\ref{rr0})--(\ref{rrk})
does not yield an appropriate asymptotic series to generate the free-energy expansion~(\ref{bif1}).
In order to regularize this critical behavior it is natural to introduce
two scaling variables $x$ and $y$ in the form
\begin{equation}
  \label{sca}
  T = 1+ \bar{\epsilon}^m x,
  \qquad
  t = 1 + \bar{\epsilon}^m y,
\end{equation}
where
\begin{equation}
  \bar{\epsilon} = \epsilon^{\frac{2}{2 m+1}}
                        =  \left(\frac{1}{N}\right)^{\frac{2}{2 m+1}}.
\end{equation}
In terms of these scaled variables the string equation~(\ref{stri22aaa}) reads
\begin{equation}
  \label{stri22bc}
  2 \bar{\epsilon}^m y {\rt} + 
  \oint_{\gamma}\frac{\rmd \lambda}{2\pi\rmi}
  V_{\lambda}(\lambda,\mathbf{g}) U(\lambda,\bar{\epsilon};{\rt})
  =
  1+\bar{\epsilon}^m x,
\end{equation}
and we will prove now that there are solutions of the form
\begin{equation}
  \label{con1}
  {\rt}(\bar{\epsilon},x,y,\mathbf{g})
  =
  r_c + \sum_{k\geq 1} {\rt}^{[k]}(x,y,\mathbf{g}) \bar{\epsilon}^k.
\end{equation}
Note that the shifts $T\rightarrow T\pm \epsilon$ correspond to $x\rightarrow x\pm \bar{\epsilon}^{1/2}$,
and therefore $U(\lambda,\bar{\epsilon};{\rt})$ is determined by the quadratic equation
\begin{equation}
  \label{resca}
  {\rt} \left(U+U_{[\overline{-1}]}\right)\left(U+U_{[\bar{1}]}\right) = \lambda\left(U^2-1\right),
\end{equation}
where we have denoted $f_{[\bar{k}]}(x)=f(x+k \bar{\epsilon}^{1/2})$.

The corresponding expansion of $U$ to be substituted in the string equation~(\ref{stri22bc}) is
\begin{equation}
  \label{stra}
  U(\lambda,\bar{\epsilon})
  =
  \sum_{k\geq 0}U^{[k]}(\lambda;{\rt}^{[1]},\ldots,{\rt}^{[k]})\, \bar{\epsilon}^k
 \end{equation}
where
\begin{equation}
  \label{uoc}
  U^{[0]} = \sqrt{\frac{\lambda}{\lambda-4 r_c}},
\end{equation}
\begin{equation}
  \label{strab}
  U^{[k]} =
  U^{[0]} \sum_{j= 1}^k \frac{ U^{[k,j]}({\rt}^{[1]},\ldots,{\rt}^{[k-j+1]})}{(\lambda-4r_c)^j}.
\end{equation}
From~(\ref{resca}) it follows that the $U^{[k,j]}$ are polynomials in
${\rt}^{[1]},\ldots,{\rt}^{[k-j+1]}$ and their $x$ derivatives, which can be determined recursively~\cite{MA07}.
In particular
\begin{equation}
 U^{[k,1]}= 2 \rt^{[k]}.
\end{equation}
The first few of these coefficients are
\begin{eqnarray}
   U^{[2,2]} & = & 6 (\rt^{[1]})^2 + 2 r_c \rt^{[1]}_{xx},\\
  U^{[3,2]} & = & 12 \rt^{[1]} \rt^{[2]} +  2\rt^{[1]}\rt^{[1]}_{xx} + 2r_c \rt^{[2]}_{xx}+ \frac{1}{6} r_c \rt^{[1]}_{xxxx},\\
  U^{[3,3]} & = & 20 (\rt^{[1]})^3  + 10 r_c (\rt^{[1]}_x)^2 + 20 r_c \rt^{[1]} \rt^{[1]}_{xx}+ 2 r_c^2 \rt^{[1]}_{xxxx}.
\end{eqnarray}

As a consequence of the quadratic equation we find the linear equation
\begin{equation}
  \label{linsa}
  {\rt}_{[\bar{1}]} \left(U_{[\bar{2}]}+U_{[\bar{1}]}\right)
  -
  {\rt} \left(U+U_{[\overline{-1}]}\right)
  =
  \lambda \left(U_{[\bar{1}]}-U\right),
\end{equation}
which in turn leads immediately to the recursion relation
\begin{equation}
  \label{pain}
  \partial_x U^{[k+1,k+1]}
  =
  \left(r_c\partial_x^3+4{\rt}^{[1]}\partial_x+2{\rt}^{[1]}_x\right)U^{[k,k]},\quad U^{[0,0]}=1.
\end{equation}
This relation implies that the $U^{[k,k]}({\rt}^{[1]})$ are the well-known
Gel'fand-Dikii differential polynomials of the KdV theory~\cite{GE75}.

We now substitute~(\ref{stra}) into~(\ref{stri22bc}) and take into account~(\ref{hj}), (\ref{mcr1}) and~(\ref{mcr2})
to obtain
\begin{equation}
  \label{straab}
  2\bar{\epsilon}^m y {\rt} + \sum_{k\geq m} W_k(r_c,\mathbf{g}) U^{[k]}(\bar{\epsilon},{\rt})
  =
  \bar{\epsilon}^m x.
\end{equation}
Finally, collecting powers of $\bar{\epsilon}$ we find the system of equations:
\begin{equation}
  \label{p}
  W_m(r_c,\mathbf{g}) U^{[m,m]}({\rt}^{[1]}) = x - 2 r_c y,
\end{equation}
\begin{equation}
  \label{mp}
  2 y {\rt}^{[k]} + \sum_{j=m}^{m+k}
                          W_j(r_c,\mathbf{g}) U^{[j,m+k]}({\rt}^{[1]},\ldots,{\rt}^{[m+k-j+1]})
  = 0,\quad k\geq 1.
\end{equation}
The first equation constrains ${\rt}^{[1]}(x,y,\mathbf{g})$ to be of the form $u(x-2 r_c y)$,
with $u(x)$ being a solution of the $m$-th member of the Painlev\'e~I hierarchy.
The subsequent equations~(\ref{mp}) give for each coefficient ${\rt}^{[k]}(x,y)$ with $(k\geq 2)$
an ordinary differential equation in the $x$ variable involving the previous coefficients $ {\rt}^{[j]}$, $(1\leq j <k)$.
The characterization of the appropriate solutions of these ordinary differential equations is a difficult problem
deeply connected to the regularization of the free energy expansion~\cite{BL05}.

To regularize the expression~(\ref{bif1}) we partition the domain $[1,+\infty)$ of the $t$ variable
into an inner region $[1,1+\delta(\bar{\epsilon})]$ and an outer region $[1+\delta(\bar{\epsilon}),+\infty)$.
In the inner region we assume the triple-scaling limit asymptotics~(\ref{con1}) for the recurrence
coefficient, while we assume the regular one-cut asymptotics~(\ref{lr}) in the outer region.
Thus, we have
\begin{equation}
  \label{bif12}
  F_{N}(\mathbf{g}) - F_N^{\mathrm{G}}
  \sim
  I_1(\bar{\epsilon},\mathbf{g}) + I_2(\epsilon,\mathbf{g}),\quad N\rightarrow\infty,
\end{equation}
where
\begin{equation}
  \label{I1}
  I_1(\bar{\epsilon},\mathbf{g})
  =
  -\bar{\epsilon}^{2 m} \int_0^{\delta(\bar{\epsilon})/\bar{\epsilon}^m}
                                   y f^{(1)}(\bar{\epsilon},y,\mathbf{g}) \rmd y,
\end{equation}
\begin{equation}
  f^{(1)}(\bar{\epsilon},y,\mathbf{g})
  =
  {\rt}(\bar{\epsilon},0,y,\mathbf{g})
  \left( {\rt}_{[\overline{-1}]}(\bar{\epsilon},0,y,\mathbf{g})
          +
         {\rt}_{[\overline{1}]}(\bar{\epsilon},0,y,\mathbf{g}) \right)
  - \frac{1}{2(1+\bar{\epsilon}^m y)^2},
\end{equation}
and
\begin{equation}
  \label{I2}
  I_2(\epsilon,\mathbf{g})
  =
  \int_{1+\delta(\bar{\epsilon})}^{\infty} (1-t) f^{(2)}(\epsilon,1,t,\mathbf{g}) \rmd t,
\end{equation}
\begin{equation}
  f^{(2)}(\epsilon,t,\mathbf{g})
  =
  {\rt}(\epsilon,1,t,\mathbf{g})\left({\rt}_{[-1]}(\epsilon,1,t,\mathbf{g})
  +
  {\rt}_{[1]}(\epsilon,1,t,\mathbf{g})\right)
  -\frac{1}{2 t^2}.
\end{equation}
To ensure that the result is independent of the choice of $\delta(\bar{\epsilon})$, 
the asymptotic series $f^{(1)}(\bar{\epsilon},y,\mathbf{g})$ and $f^{(2)}(\epsilon,t,\mathbf{g})$
must be matched on some appropriate intermediate region overlapping the inner and outer
regions. It is at this point where the conditions to determine the coefficients of~(\ref{con1}) emerge.
\subsection{The Brezin-Marinari-Parisi critical model}
We illustrate these ideas with the Brezin-Marinari-Parisi critical potential~(\ref{bmp}), which
belongs to the singular one-cut case under the Bleher-Its deformation.  In this case
\begin{equation}
W({\rt}_0,\mathbf{g})={\rt}_0^3 -3 
{\rt}_0^2+3 {\rt}_0,
\end{equation}
where $\mathbf{g}=(g_2,g_4,g_6)=(3/2,-1/4,1/60)$, and according to~(\ref{mcr1})--(\ref{mcr2})
we have a critical point of order $m=3$ at $r_c=1$. Thus, from~(\ref{p}) it follows that 
$ {\rt}^{[1]}=u(x-2 y)$, where $u(x)$ is a solution of the second member of the Painlev\'e~I
hierarchy
\begin{equation}
  \label{p14}
  u_{xxxx}+10 u u_{xx} + 5 u_x^2 + 10 u^3 = 10 x.
\end{equation}

To perform the matching between the triple-scaling and the one-cut regular asymptotics,
we note that as $t\rightarrow 1^+$, from~(\ref{rr0}) and~(\ref{rrk}) we have
\begin{equation}
  {\rt}_0(1,t,\mathbf{g}) \sim 1-2^{1/3} (t-1)^{\frac{1}{3}},
\end{equation}
\begin{equation}
  {\rt}''_0(1,t,\mathbf{g}) \sim 3^{-2} 2^{-2/3} (t-1)^{-\frac{5}{3}},
\end{equation}
\begin{equation}
  {\rt}_1(1,t,\mathbf{g}) \sim \frac{1}{72 (t-1)^2}.
\end{equation}
Hence we get
\begin{equation}
  \label{asee}
  f^{(2)}(\epsilon,t,\mathbf{g})
  \sim \left( 2 {\rt}_0^2 - \frac{1}{2 t^2} \right) + \epsilon^2 {\rt}_0 (4 {\rt}_1+ {\rt}_0'')
  \sim \frac{3}{2} - \bar{\epsilon} 2^{7/3}y^{1/3}.
\end{equation}
Likewise, in the inner region we have
\begin{equation}
  \label{f1i}
  f^{(1)}(\bar{\epsilon},y,\mathbf{g})
  \sim
  \frac{3}{2} + \bar{\epsilon} 4 {\rt}^{[1]}(0,y,\mathbf{g}).
\end{equation}
In the matching region we must have both $t\rightarrow 1^+$ and
$y\rightarrow +\infty$ as $N\rightarrow\infty$. Therefore, the matching between~(\ref{asee})
and~(\ref{f1i}) is achieved provided $u(x)$ is a solution of~(\ref{p14}) such that
\begin{equation}
  \label{cv}
  u(-2 y)\sim -2^{1/3} y^{1/3}, \quad y\rightarrow +\infty.
\end{equation}
This asymptotic behavior determines a unique formal expansion of the form $x^{1/3}$ times a
series in powers of $x^{-7/3}$, which solves~(\ref{p14}).
\section{Concluding remarks}
In this paper we have developed a method  to compute the  large $N$ expansion of  the free energy 
of Hermitian matrix models~(\ref{bif20}) from the  large $N$ expansion of the recurrence coefficients
of the associated family of orthogonal polynomials. It is based on  the Bleher-Its deformation, on its associated
integral representation of the free energy, and on a method for solving the string equation which
uses the resolvent of the Lax operator of the underlying Toda hierarchy.  Combining these ingredients we
provide a  procedure,  suitable for symbolic computation, to characterize the structure of the coefficients
$F^{(k)}(\mathbf{g})$  of the topological expansion of the free energy.  The procedure  can be also used
efficiently  for the explicit evaluation of these coefficients.  As an illustrative application  we compute
the expressions of  $F^{(k)}(\mathbf{g})\, (k=0,\ldots,3)$  for general matrix models and check their
agreement  with the expressions derived using  the Euler-Maclaurin summation formula in the
Bessis-Itzykson-Zuber method. 

The main application of our study is a convenient method to  compute  generating functions for the
enumeration of labeled $k$-maps. It relies on the structure of the integral representations of the
coefficients $F^{(k)}(\mathbf{g})$ and does not require the explicit expressions of these  coefficients.
We apply this method to elaborate several tables of numbers of   $k$-maps with two and three valences
and up to genus $k=4$. 

Finally, in order to illustrate the regularization of singular models within our scheme we have
formulated a triple-scaling method to regularize singular one-cut models. 

Although in this paper we restricted our analysis to the genus expansions of one-cut even models,
since both the Bleher-Its representation of the free-energy~\cite{BL05} and the method for solving
the string equation using the resolvent of the Lax operator~\cite{MA07} can be used for models
associated with general (not necessarily even) potentials, there is no obstacle to apply
our analysis to these problems too. Furthermore, Bleher and Its~\cite{BL05} applied the integral
representation~(\ref{bif}) to determine a three-term large $N$ asymptotic expansion for the free
energy for the quartic model in the neighborhood of a critical point at the boundary between
the phases $G_1$ and $G_2$. The third of these terms, which involves the
Tracy-Widom distribution function~\cite{TR94}, represents a nonperturbative effect.
Thus, it is  plausible to generalize the present scheme to characterize nonperturbative
contributions to the large $N$ asymptotics of the free energy of multi-cut models.
This generalization would require the use of nonperturbative solutions of the string equation, e.g.,
as in the trans-series method considered by Mari\~no~\cite{MA08b}.
In particular it would be interesting to investigate the asymptotic behavior of the free energy
in critical processes such as the birth of a cut in the eigenvalue support~\cite{EY06,FL08,AL10}.
\section*{Acknowledgments}
The financial support of the Universidad Complutense under project GR58/08-910556 and the Comisi\'on
Interministerial de Ciencia y Tecnolog\'{\i}a under projects FIS2008-00200 and FIS2008-00209 are gratefully acknowledged.
\section*{Appendix A: Multi-cut models}
We first recall the following upper bound~\cite{AL10} for the number of cuts $q$ of a model with the potential $V(z)$:
\begin{equation}
  \label{nice}
  q \leq p = \frac{\mbox{deg} V(z)}{2}.
\end{equation}
The conditions that determine the actual value of $q$ among those allowed by this bound can be stated in terms of the
function $h(z)$ defined in~(\ref{0.3}). It can be shown~\cite{DE99b} that in the  $q$-cut case:
\begin{enumerate}
\item The endpoints  of $J$  satisfy the equations
\begin{equation}
  \label{e1}
  \int_{\beta_{j}}^{\alpha_{j+1}}h(x) w_{1,+}(x) \rmd x = 0,\quad j=1,\ldots,q-1,
\end{equation}
\begin{equation}
  \label{e2}
  \oint_{\gamma}z^j \frac{V_z(z)}{w_1(z)} \rmd z = 0,\quad j=0,\ldots,q-1,
\end{equation}
where $\gamma$ is a large positively oriented loop around $J$. Moreover, since $\int_J \rho(x) \rmd x=1$ we must have
\begin{equation}
  \label{e3}
  \oint_{\gamma}h(z) w_{1}(z) \rmd z = -4\pi\rmi .
\end{equation}
\item The following inequalities hold:
\begin{equation}
  \label{des1}
  \int_x^{\alpha_1} h(x') w_{1}(x') \rmd x'\leq 0,\quad \mbox{for $x<\alpha_1$} ,
\end{equation}
\begin{equation}
  \label{des2}
  \int_{\beta_{j}}^x h(x') w_{1}(x') \rmd x'\geq 0,\quad \mbox{for $\beta_{j}<x<\alpha_{j+1},\quad j=1,\ldots q-1$},
\end{equation}
\begin{equation}
  \label{des3}
  \int_{\beta_{q}}^x h(x') w_{1}(x') \rmd x'\geq 0,\quad \mbox{for $x>\beta_{q}$}.
\end{equation}
\end{enumerate}
Equations~(\ref{e1})--(\ref{e3}) are $2 q$ conditions that the $2 q$ unknowns $\alpha_1,\ldots,\beta_q$
must satisfy. However, these equations may not be sufficient to determine uniquely $q$ because they may have
admissible solutions for different values of $q$. If this is the case, the additional condition $\rho(x)>0$ for all $x\in J$
and the inequalities~(\ref{des1})--(\ref{des3}) characterize uniquely the solution of the problem.
A  model is said to be a \emph{regular}  if $h(x)\neq 0$ on $\bar{J}$ and the
inequalities~(\ref{des1})--(\ref{des3}) are strict. Otherwise it is called \emph{singular}.
\subsection*{Sixtic potentials}
Let us consider the family of sixtic potentials
\begin{equation}
  V(\lambda) = g_2 \lambda + g_4 \lambda^2 + g_6 \lambda^3,\quad (\lambda=z^2),
\end{equation}
in the region of coupling constants
\begin{equation}
  G=\{\mathbf{g} =(g_2,g_4,g_6)\in \mathbf{R}^3:  g_2>0, g_4<0,  g_6>0\}.
\end{equation}
For $q=1$ and  $J=(-\alpha,\alpha)$ we have
 \begin{equation}
  w_{1,+}(x) = \left\{ \begin{array}{ll}
                               - |x^2-\alpha^2|^{1/2} & \mbox{ for $x\leq - \alpha$},\\
                               i |x^2-\alpha^2|^{1/2} & \mbox{ for $-\alpha \leq x \leq \alpha$},\\
                                |x^2-\alpha^2|^{1/2}  & \mbox{ for $x\geq \alpha$},
                              \end{array}\right.
\end{equation}
and
\begin{equation}
h(x)=6 g_6  x^4+(4 g_4 +3 g_6 \alpha^2) x^2+ \frac{9}{4} g_6 \alpha^4+2 g_4 \alpha^2+2 g_2.
\end{equation}
Equation~(\ref{q1}) reads
\begin{equation}
  \label{q11}
  15 g_6 A^3 + 12 g_4 A^2 + 8 g_2 A - 16 = 0,\quad (A=\alpha^2).
\end{equation}
Completing  squares in the expression of $h(x)$ we have
\begin{equation}
  \label{sqr}
  h(x) = 6 g_6 \left(x^2+\frac{\alpha^2}{4}+\frac{g_4}{3 g_6} \right)^2
            +
            \frac{15 g_6}{8} \left(\alpha^2+ \frac{4 g_4}{15 g_6} \right)^2
            +
            2 g_2 - \frac{4 g_4^2}{5  g_6}.
\end{equation}
Hence the function $h(x)$ is strictly positive for all $x\in\mathbf{R}$ provided that
\begin{equation}
  \label{inn}
  \frac{5}{2}\frac{g_2 g_6}{g_4^2}>1.
\end{equation}
Thus, the inequalities~(\ref{des1})--(\ref{des3}) are  strictly satisfied and $\rho(x)>0$ on $J$.
Moreover, since the critical points of the polynomial in the left-hand side of~(\ref{q11}) are
\begin{equation}
  A = \frac{4 |g_4|}{15 g_6} \left(1\pm\sqrt{1-\frac{5}{2} \frac{g_2  g_6}{g_4^2}}\right),
\end{equation}
then~(\ref{inn}) implies that there exists a unique positive solution $A$ of~(\ref{q11}).
Therefore~(\ref{inn}) determines an open subset of $G_1$ with boundary $\Gamma$  given by
\begin{equation}
  \label{innb}
  2 g_4^2 = 5 g_2 g_6.
\end{equation}
Given $\mathbf{g}\in \Gamma$ it follows from~(\ref{sqr}) that  the function $h(x)$ is strictly
positive on $\mathbf{R}$ unless
\begin{equation}
  \label{bb}
  \alpha^2 = -\frac{4 g_4}{15 g_6},
\end{equation}
in which $h(x)$ vanishes at $x=\pm \alpha$.  But this value of $\alpha^2$ satisfies
equation~(\ref{q11}) only if
\begin{equation}
  \label{hb}
  4 g_4^3 = -225 g_6^2.
\end{equation}
Hence, along the curve $\gamma$ given by
\begin{equation}
  \label{cur}
  2 g_4^2 = 5 g_2 g_6,\quad 4 g_4^3 = -225 g_6^2,
\end{equation}
the model is singular because $h(x)$ vanishes at the end-points $\pm \alpha$ of
the eigenvalue support, whereas for points in $\Gamma-\gamma$ the model is in $G_1$.
\section*{Appendix B: The quadratic equation for $U_{n,N}$}
For clarity in this appendix we drop the subindex $N$ from $U_{n,N}$. Thus, consider the function
\begin{equation}
  \label{gen1}
  U_n(\lambda) = 1 + 2 \sum_{k\geq 1} (L^{2k-1})_{n,n-1} \lambda^{-k},
\end{equation}
where the matrix elements $(L^{2k-1})_{n,n-1}$ are calculated in the basis of orthogonal polynomials
\begin{equation}
  v_n(x) = P_{n,N}(x),\quad n\geq 0.
\end{equation}
From~(\ref{rec}) it follows that $L^{2k-1} v_n(x)=x^{2k-1} v_n(x)$, and therefore
\begin{equation}
  (L^{2k-1})_{n,n-1}
  =
  \frac{1}{h_{n-1,N}}
  \int_{-\infty}^{\infty} x^{2k-1} v_n(x) v_{n-1}(x) \rmd \mu(x),
\end{equation}
where
\begin{equation}
  \rmd \mu(x) = \rme^{-N V(x)} \rmd x.
\end{equation}
Hence,
\begin{equation}
  \label{expu}
  U_n = 1+\frac{2}{h_{n-1,N}}
             \int_{-\infty}^{\infty} \frac{x v_n(x) v_{n-1}(x)}{\lambda-x^{2}}\rmd \mu(x).
\end{equation}

Let us  prove that $U_n$ satisfies the linear equation
\begin{equation}
  \label{lin}
  \lambda (U_{n+1}-U_n) = r_{n+1,N} (U_{n+2}+U_{n+1}) - r_{n,N} (U_{n}+U_{n-1}).
\end{equation}
From~(\ref{expu}) we deduce that
\begin{eqnarray}
  \label{expu1}
  \lambda (U_{n+1}-U_n)
  & = & \frac{2}{h_{n,N}} \int_{-\infty}^{\infty} x v_{n+1}(x) v_{n}(x) \rmd \mu(x)\nonumber\\
  &    & {}-\frac{2}{h_{n-1,N}} \int_{-\infty}^{\infty} x v_{n}(x) v_{n-1}(x) \rmd \mu(x)\nonumber\\
  &    & {}+\frac{2}{h_{n,N}} \int_{-\infty}^{\infty} \frac{x^3 v_{n+1}(x) v_{n}(x)}{\lambda-x^{2}}\rmd \mu(x)\nonumber\\
  &    & {}-\frac{2}{h_{n-1,N}} \int_{-\infty}^{\infty} \frac{x^3 v_n(x) v_{n-1}(x)}{\lambda-x^{2}}\rmd \mu(x).
\end{eqnarray}
Using $x^j v_n(x) =L^j v_n(x)$ for $j=1,2$ we find that
\begin{equation}
  \int_{-\infty}^{\infty} x v_{n+1}(x) v_{n}(x) \rmd \mu(x) = h_{n+1,N},
\end{equation}
and
\begin{eqnarray}
 && \int_{-\infty}^{\infty} \frac{x^3 v_{n+1}(x) v_{n}(x)}{\lambda-x^{2}}\rmd \mu(x)  = \nonumber\\
 &&\int_{-\infty}^{\infty}
  \frac{x v_{n+2} v_{n+1} +x (r_{n+1,N}+r_{n,N}) v_n v_{n+1} + x r_{n,N}r_{n-1,N} v_{n+1} v_{n-2}}{
  \lambda-x^{2}}\rmd\mu(x).\nonumber\\
  &&
\end{eqnarray}
Substituting these identities in~(\ref{expu1}) and taking into account~(\ref{expu}) we conclude that~(\ref{lin}) holds.

It is now easy to prove that
\begin{equation}
  \label{res12}
  r_{n,N} (U_n+U_{n-1}) (U_{n}+U_{n+1}) = \lambda (U_n^2-1).
\end{equation}
Indeed, the linear identity~(\ref{lin}) implies
\begin{equation}
  r_{n+1,N} (U_{n+2}+U_{n+1}) (U_{n+1}+U_{n})
  -
  r_{n,N} (U_{n+1}+U_{n}) (U_{n}+U_{n-1})
  =
  \lambda (U_{n+1}^2-U_{n}^2),
\end{equation}
and therefore the expression
\begin{equation}
 \lambda U_{n}^2-r_{n,N} (U_{n+1}+U_{n}) (U_{n}+U_{n-1})
\end{equation}
is independent of $n$. Since $r_{0,N}=0$ and $U_0=1$ the identity~(\ref{res12}) follows.

\end{document}